\documentclass[12pt]{article}
\usepackage{graphicx}
\usepackage{subfigure}
\usepackage{graphicx,epsfig,amssymb}
\usepackage{epstopdf}
\usepackage{setspace}
\usepackage{amsmath}
\usepackage{mathtools}
\usepackage{amsfonts,amssymb}
\usepackage{subfig}
\usepackage{placeins}
\usepackage{color}
\usepackage[percent]{overpic}

\def\l {\label}

\def\be {\begin{equation}}
\def\ee {\end{equation}}
\def\ba {\begin{eqnarray}}
\def\ea {\end{eqnarray}}
\newcommand{\dss}{\displaystyle}

\begin{document}
\baselineskip=15pt
\renewcommand {\thefootnote}{\dag}
\renewcommand {\thefootnote}{\ddag}
\renewcommand {\thefootnote}{ }

\pagestyle{empty}

\begin{center}
                \leftline{}
                \vspace{-0.50 in}

{\Large \bf Visualization of Four Limit Cycles in \\ [1.0ex] 
Near-Integrable Quadratic Polynomial Systems} \\ [0.3in]

{\large Pei\, Yu$^*$, \ Yanni Zeng}
\footnote{$^*\,$Corresponding author, Email: pyu@uwo.ca}
\\ 

\vspace{0.15in}
{\small {\em
Department of Applied Mathematics, Western University \\[-0.5ex]
London, Ontario, Canada \ N6A 5B7
}}


\end{center}

\baselineskip=15pt 
\vspace{0.20in}

\noindent 
\rule{6.5in}{0.012in}

\vspace{-0.2in}
\section*{Abstract}
It has been known for almost $40$ years that general planar quadratic 
polynomial systems can have four limit cycles. Recently, four limit cycles 
were also found in near-integrable quadratic polynomial systems. 
To help more people to understand limit cycles theory,   
the visualization of such four numerically simulated 
limit cycles in quadratic systems has attracted researchers' attention. 
However, for near-integral systems, such visualization becomes much more 
difficult due to limitation on choosing parameter values.  
In this paper, we start from the simulation of the well-known quadratic systems 
constructed around the end of 1979, then  
reconsider the simulation of a recently published quadratic system 
which exhibits four big size limit cycles, and finally provide 
a concrete near-integral 
quadratic polynomial system to show four normal size limit cycles.

\vspace{0.1in} 

\noindent 
{\it Keywords}: Hilbert's 16th problem, quadratic near-integrable system, 
limit cycle, 

\hspace{0.50in}Hopf bifurcation, Poincar\'{e} bifurcation, Melnikov function  

\vspace{0.10in} 
\noindent
{\it MSC}: 34C07; 34C23 

\noindent 
\rule{6.5in}{0.012in}
\vspace{0.05in}

\section{Introduction}\label{intro}

The well-known Hilbert's 16th problem is remained unsolved 
for more than one hundred years since Hilbert~\cite{hilbert1900} 
proposed the $23$ mathematical problems. 
A simplified version of the problem, based on a general 
Li\'{e}nard equation, was chosen by Smale~\cite{smale1998} as 
one of the $18$ challenging mathematical problems for the $21$st
century. Consider the following planar system: 
\begin{eqnarray}
\dot{x} = P(x,y), \qquad \dot{y} = Q(x,y),
\label{b1}
\end{eqnarray}
where the dot denotes differentiation with respect to time $t$, 
$P(x,y) $ and $Q(x,y) $ are 
polynomials in $x$ and $y$. The second part of Hilbert's $16$th 
problem is to find the upper bound, called Hilbert number 
and denoted by $H(n)$, where $n=\max\{{\rm deg}P,\, {\rm deg}Q\}$, 
on the number of limit cycles that system 
\eqref{b1} can have. 
If the problem is restricted to
the neighborhood of isolated fixed points, then the
question is reduced to studying degenerate Hopf bifurcations. 
In 1952, Bautin~\cite{bautin} proved that three small limit cycles exist  
around a fine focus or a center in quadratic systems. Almost 30 years later,
concrete examples were independently constructed by 
Shi~\cite{Shi79}, and by Chen and Wang~\cite{ChenWang}  
to show the existence of $4$ limit cycles in quadratic 
systems, implying that $H(2) \ge 4$. 
However, the question whether $H(2)=4$ is still open. 

\pagestyle{myheadings}
\markright{{\footnotesize {\it P. Yu and Y. Zeng} 
\hspace{1.2in} {\it Visualization of $4$ limit cycles in quadratic systems}}}

To reduce the difficulty in attacking the Hilbert's 16th problem, 
Arnold proposed a weak version of the problem \cite{Arnold1977}, 
which transforms the problem of determining the maximal number of limit 
cycles (a geometric problem) to finding the maximal number of isolated 
zeros of the Abelian integral or Melnikov function (an algebraic problem): 
\be 
M(h) = \dss\oint_{H(x,y)=h} Q(x,y) \, dx - P(x,y) \, dy, 
\l{b2} 
\ee 
where $ H(x,y), \, P $ and $ Q $ are all real polynomials 
in $x $ and $ y$ with $ {\rm deg} H = n+1$, 
and $ \max\{ {\rm deg} P, \, {\rm deg} Q \} \le n$. 
The weak Hilbert's 16th problem is closely 
related to the maximal number of limit cycles in the following 
near-Hamiltonian system~\cite{Han2006}: 
\be 
\begin{array}{ll}  
\dot{x} = \dss\frac{\partial H(x,y)}{\partial y}
+ \varepsilon \, p_n(x,\,y), \\[2.0ex] 
\dot{y} =- \,  \dss\frac{\partial H(x,y)}{\partial x} 
+ \varepsilon \, q_n(x,\,y), 
\end{array}
\l{b3} 
\ee
where $ p_n (x,y) $ and $ q_n(x,y) $ are $n$th-degree polynomials 
in $x$ and $y$, and $ H(x,y)$ is a polynomial in $x$ and $y$ 
with ${\rm deg}H=n+1$, and $ 0 < \varepsilon \ll 1 $ 
indicating the small perturbations 
$\varepsilon \, p_n(x,\,y)$ and $\varepsilon \, q_n(x,\,y)$ 
on the system. When $\varepsilon=0$, \eqref{b3} 
is reduced to a Hamiltonian system, and the first-order ($\varepsilon$-order) 
Melnikov function becomes 
\be 
M(h) = \dss\oint_{H(x,y)=h} q_n(x,y) \, dx - p_n(x,y) \, dy. 
\l{b4} 
\ee

In this paper, we focus on quadratic systems, i.e., on the case 
$n=2$ in (\ref{b1}), and pay particular attention to the numerical 
realization of four limit cycles which are visualizable. 
In general, a quadratic system which has at least two singularities 
can be written in the form of~\cite{ChenWang,Ye1982} 
\be  
\begin{array}{ll} 
\dot{x} = -y + l\, x^2 + m\, xy + n\, y^2, \\ [1.0ex]  
\dot{y} = x ( 1 + a\, x + b\, y), 
\end{array}
\l{b5} 
\ee 
which, under the assumption $n\ne 0$, can be rescaled to 
\be  
\begin{array}{ll} 
\dot{x} = -y + l\, x^2 + m\, xy + y^2, \\ [1.0ex]  
\dot{y} = x\, ( 1 + a\, x + b\, y). 
\end{array}
\l{b6} 
\ee 
System \eqref{b6} is exactly the same as the system given in~\cite{YuHan2012} 
under the transformation $(x,y) \rightarrow (y,x)$: 
\be  
\begin{array}{ll} 
\dot{x} = y\,(1 + a_1\, x + a_2\, y), \\ [1.0ex]  
\dot{y} = -\, x + x^2 + a_3\, x\, y + a_4\, y^2. 
\end{array}
\l{b7} 
\ee
Note that system \eqref{b6} has two singularities at 
$(0,0)$ and $(0,1)$, while system \eqref{b7} has 
two singularities at $(0,0)$ and $(1,0)$. 

Recently, Kuznetsov {\it et al.}~\cite{KKL2013} 
considered the following quadratic system: 
\be  
\begin{array}{ll} 
\dot{x} = y + x^2 + x\, y , \\ [1.0ex]  
\dot{y} = a_2 \, x^2 + b_2 \, x\, y + c_2 \, y^2 + \alpha_2 \, x 
+ \beta_2 \, y, 
\end{array}
\l{b8} 
\ee
and proved that the system has four limit cycles if certain 
conditions on the parameters are satisfied. In particular, they 
chose a set of parameter values to show four big size limit cycles. 

To consider perturbing an integrable system, we need 
$(0,0)$ to be a center, for which it has the classifications 
as follows. 
The origin of (\ref{b7}) is a center if and only 
if one of the following conditions is satisfied~\cite{YuHan2012}, 
$$
\begin{array}{ll} 
Q_3^{\rm R} \!\!\!\! & - \ \ \textrm{Reversible system}: 
\ a_3 = a_2 = 0; \\[1.0ex]  
Q_3^{\rm H} \!\!\!\! & - \ \ \textrm{Hamiltonian system}: 
\ a_3 = a_1 + 2\, a_4 = 0; \\[1.0ex] 
Q_3^{\rm LV} \!\!\!\! & - \ \ \textrm{Lokta-Volterra system}: 
\ a_2 = 1+a_4 = 0; \ \textrm{and} \\[1.0ex] 
Q_4 \!\!\!\! & - \ \ \textrm{Codimension-4 system}: 
\ a_3 - 5\, a_2 = a_1 - (5 + 3\, a_4) = a_4 + 2 (1+a_2^2) = 0. 
\end{array} 
$$
In this paper, we will perturb the reversible system to 
obtain the following perturbed one~\cite{YuHan2012}: 
\be  
\begin{array}{rl} 
\dot{x} \!\!\! & = y \, (1 + a_1\, x) 
+ \varepsilon \, p_n(x,y) \\[1.0ex] 
& = y\, (1 + a_1\, x) + \varepsilon \, a_{10}\, x, \\ [1.0ex]  
\dot{y} \!\!\! & = -\, x + x^2 + a_4\, y^2 + \varepsilon \, q_n(x,y) \\[1.0ex]
& = -\, x + x^2 + a_4\, y^2 + \varepsilon \, (b_{01}\, y + b_{11}\, x\, y), 
\end{array}
\l{b9} 
\ee 
where $a_{10}, \, b_{01}$ and $b_{11}$ are perturbation parameters. 

In this paper, we consider bifurcation of limit cycles in the 
quadratic systems \eqref{b6}, \eqref{b8} and \eqref{b9}, and show 
visualizable simulated four limit cycles.  
Since the convergence for small limit cycles are extremely slow, 
particularly for system \eqref{b9},  
we apply the Runge-Kutta (R-K) $4$th-order method to simulate 
small limit cycles 
on a Desktop machine with CPU@$3.20$GHz, and use Matlab solver ODE23 to 
simulate large limit cycles. For unstable limit cycles, we use 
backward time (i.e., using negative time step) and so the unstable limit 
cycles become ``stable''.  
In next section, we consider the quadratic examples proposed 
by Shi~\cite{Shi79}, and by Chen and Wang~\cite{ChenWang}.  
In Section 3, we reconsider the example constructed by 
Kuznetsov {\it et al.}~\cite{KKL2013} to demonstrate four big 
size limit cycles. Finally, we present a concrete 
near-integrable quadratic system~\cite{YuHan2012}
to show four normal size limit cycles.

\section{Four limit cycles obtained in~\cite{ChenWang} and \cite{Shi79}}

In this section, we consider the two concrete quadratic systems 
given by Shi~\cite{Shi79} 
and by Chen and Wang~\cite{ChenWang}. 
The Shi's example is given by the following equations~\cite{Shi79}: 
\be  
\begin{array}{ll} 
\dot{x} = \lambda\, x -y -10\, x^2 + (5 + \delta)\, xy + y^2, \\ [1.0ex]  
\dot{y} = x + x^2 + (-25 + 8 \epsilon - 9 \delta) \, x\, y, 
\end{array}
\l{b10} 
\ee 
where the parameter values chosen in \cite{Shi79} are 
\be  
\lambda = -\,10^{-250}, \quad 
\epsilon = -\, 10^{-52}, \quad  
\delta = -\,10^{-13},
\l{b11} 
\ee
for proving the existence of four limit cycles.  
The example given by Chen and Wang in~\cite{ChenWang} is described by 
\be  
\begin{array}{ll} 
\dot{x} = -\delta_2 \, x -y -3\, x^2 + (1 - \delta_1)\, xy + y^2, \\ [1.0ex]  
\dot{y} = x + \dfrac{2}{9}\, x^2 -3\, x\, y, 
\end{array}
\l{b12} 
\ee 
where $0 < \delta_1,\, \delta_2 \ll 1$, but not specified.  

\begin{figure}[!h]
\vspace{ 0.20in} 
\hspace{0.30in} 
\includegraphics[width=0.34\textwidth,height=0.235\textheight,angle=-00]{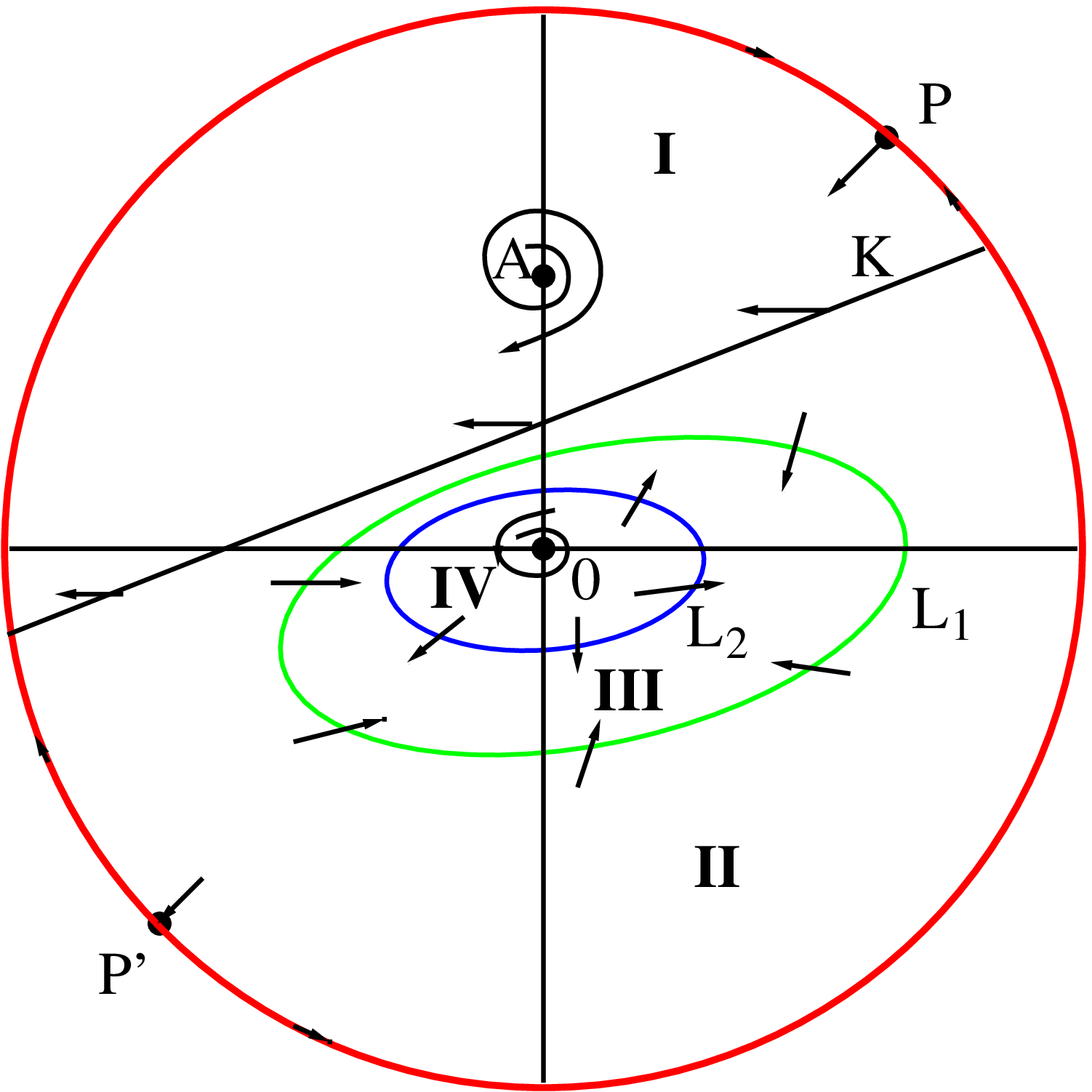}

\vspace{-2.11in} 
\hspace{ 3.50in} 
\includegraphics[width=0.34\textwidth,height=0.235\textheight,angle=-00.0]{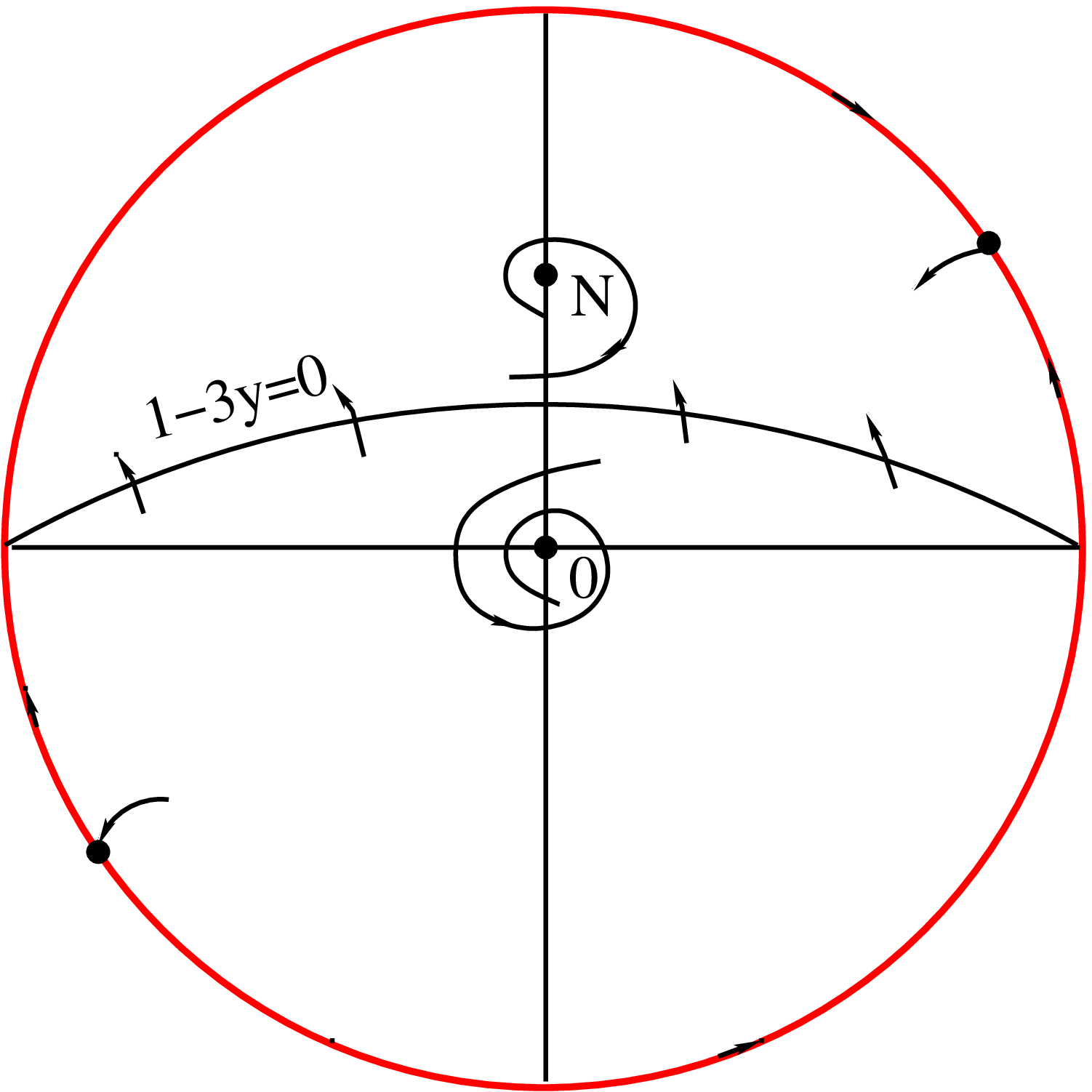}

\vspace{0.10in} 
\hspace{1.35in}(a)\hspace{3.00in}(b) 

\vspace{ 0.10in} 
\caption{Schematic diagrams showing the existence of limit cycles 
for (a) system \eqref{b10} (see Fig. $1$ in~\cite{Shi79}); and 
(b) system \eqref{b12} (see Fig. $3$ in~\cite{ChenWang}).} 
\label{fig1}
\end{figure}

Firstly it is noted that the two systems \eqref{b10} and \eqref{b12} 
have the exact same structure of \eqref{b6}. 
Secondly it was proved that both systems 
have a big stable limit cycle around the unstable focus $(0,1)$, and three 
small limit cycles around the stable focus $(0,0)$.    
The schematic diagrams showing the 
existence of limit cycles are depicted in Figure~\ref{fig1}.

\begin{figure}[!h]
\hspace{0.30in} 
\begin{overpic}[width=0.43\textwidth,height=0.256\textheight,angle=-00]{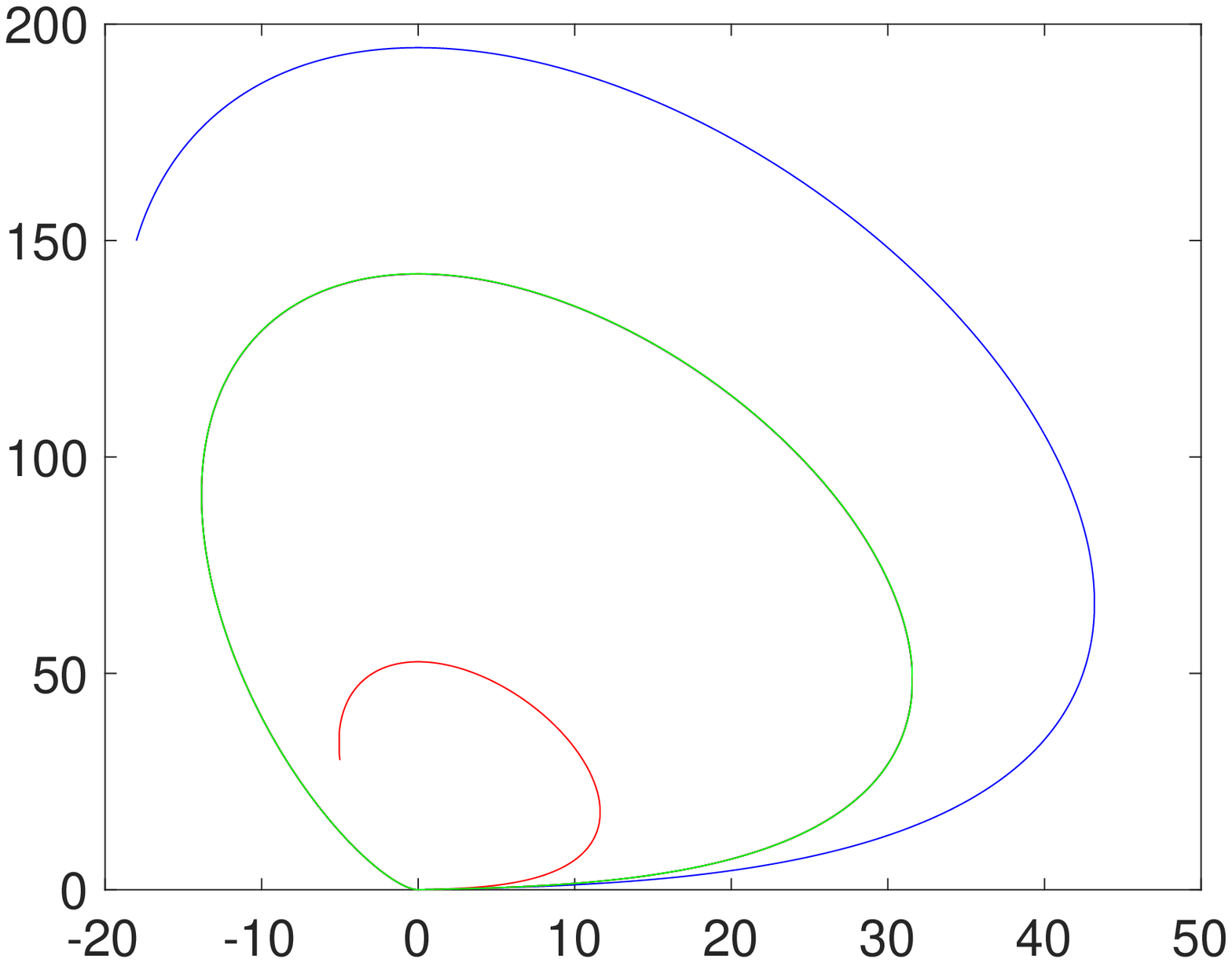} 

\put(14.1,58.0){\color{blue}{\footnotesize{$\bullet$}}}  
\put(28.4,17.8){\color{red}{\footnotesize{$\bullet$}}} 

\put(61.4,65.0){\color{blue}{\vector(1,-1){3}}} 
\put(54.4,50.0){\color{green}{\vector(1,-1){3}}} 
\put(42.4,24.0){\color{red}{\vector(1,-1){3}}} 
 
\put(79.0,67.0){\color{black}(a)}
\end{overpic} 

\vspace*{-2.31in} 
\hspace{3.30in} 
\begin{overpic}[width=0.43\textwidth,height=0.269\textheight,angle=-00]{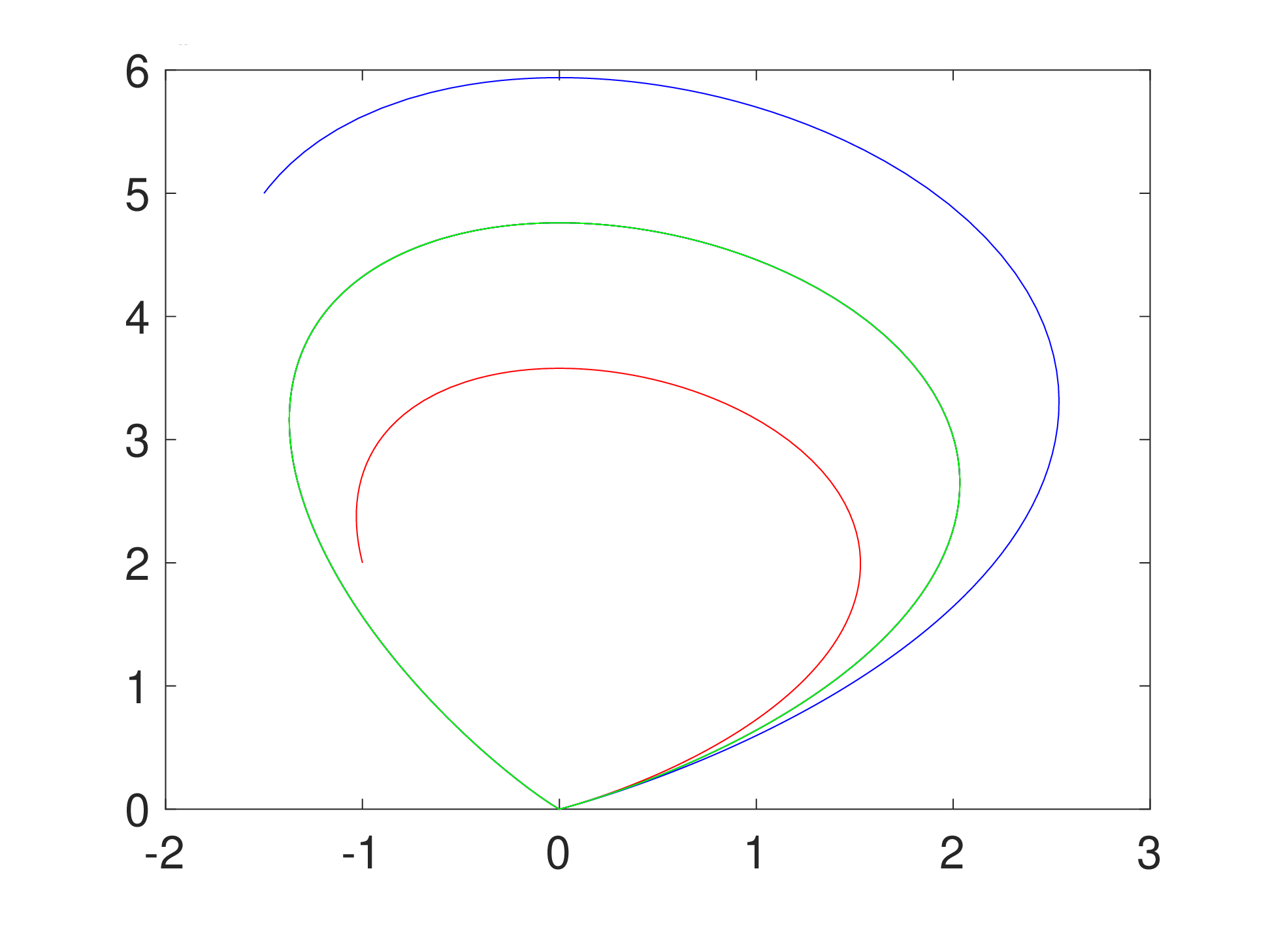}

\put(12.0,81.0){\tiny{$\times 10^{11}$} }
\put(92.0,8.2){\tiny{$\times 10^{11}$} }

\put(19.7,67.4){\color{blue}{\footnotesize{$\bullet$}}}  
\put(27.5,33.9){\color{red}{\footnotesize{$\bullet$}}}  

\put(61.4,75.8){\color{blue}{\vector(2,-1){3}}} 
\put(54.4,64.5){\color{green}{\vector(3,-1){3}}} 
\put(46.4,52.9){\color{red}{\vector(4,-1){3}}} 
 
\put(78.0,71.0){\color{black}(b)}

\end{overpic}

\vspace{-0.10in}  
\caption{Simulation of big limit cycle for  
(a) system \eqref{b10} when $\lambda=\delta=\epsilon=0$ with 
initial points $(-18,150)$ and $(-5,30)$; and  
(b) system \eqref{b12} when $\delta_1=\delta_2 =0$ 
with initial points $(-1.5 \! \times \! 10^{11},5 \! \times \! 10^{11})$ 
and $(-10^{11},2 \! \times \! 10^{11})$.} 
\label{fig2} 
\vspace{0.10in} 
\end{figure}

Note that when $\lambda = \epsilon =\delta=0$, the origin 
of system \eqref{b10} is a $3$rd-order fine focus, while 
the origin of system \eqref{b12} is a $2$nd-order fine focus  
when $\delta_1 = \delta_2 = 0$.
In \cite{Shi79} Shi explicitly constructed four trapping regions 
(see Figure~\ref{fig1}(a)) 
and applied 
Poincar\'{e}-Bendixson theory to prove the existence of four limit cycles, one 
of them around $(0,1)$ and three of them around $(0,0)$ which were 
obtained by perturbing the $3$rd-order fine focus using 
the three parameters. In \cite{ChenWang} Chen and Wang constructed 
two trapping regions (see Figure~\ref{fig1}(b)) 
and used Poincar\'{e}-Bendixson theory to prove 
the existence of the big limit cycle around $(0,1)$ and a small one 
around $(0,0)$. Then they showed that further perturbing 
the $2$nd-order fine focus $(0,0)$ using the parameters 
$\delta_1$ and $\delta_2$ to obtain two more small limit cycles, 
but did not specify the values of 
$\delta_1$ and $\delta_2$.     
The stability of the four limit cycles is same for both systems: 
The big one around the unstable focus $(0,1)$ is stable, and the most outer 
small limit cycle around $(0,0)$ is unstable, the middle one is stable and 
the most inner one is unstable. The focus $(0,0)$ is stable.  
Figure~\ref{fig2} shows the simulation of 
the big limit cycle for systems \eqref{b10} and \eqref{b12} when 
$\lambda=\epsilon = \delta=\delta_1=\delta_2=0$. Two initial points are 
chosen for simulating the trajectories with one outside the limit cycle 
(in blue color) and one inside the limit cycle (in red color), both 
converging to the stable limit cycle (in green color).

\begin{figure}[!h] 
\vspace{0.20in} 
\begin{center}
\begin{overpic}[width=0.3\textwidth,height=0.3\textheight,angle=-90]{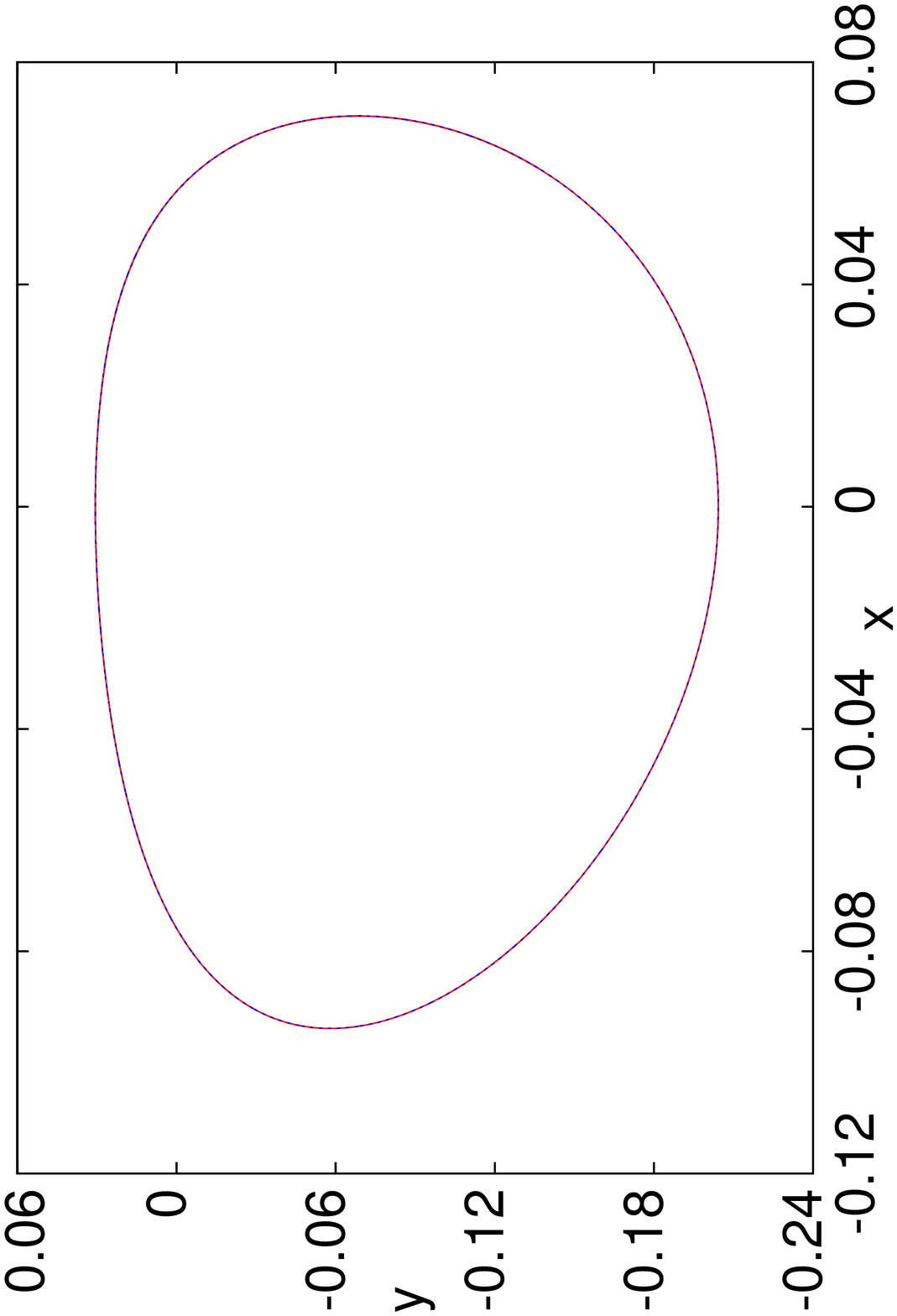}

\put(61.0,56.2){\footnotesize{$\bullet$}}  

\put(28.4,29.8){\color{black}{\vector(1,-1){3}}} 

\put(15.0,61.0){\color{black}(a)}
\end{overpic}
\hspace*{0.30in}
\begin{overpic}[width=0.3\textwidth,height=0.322\textheight,angle=-90]{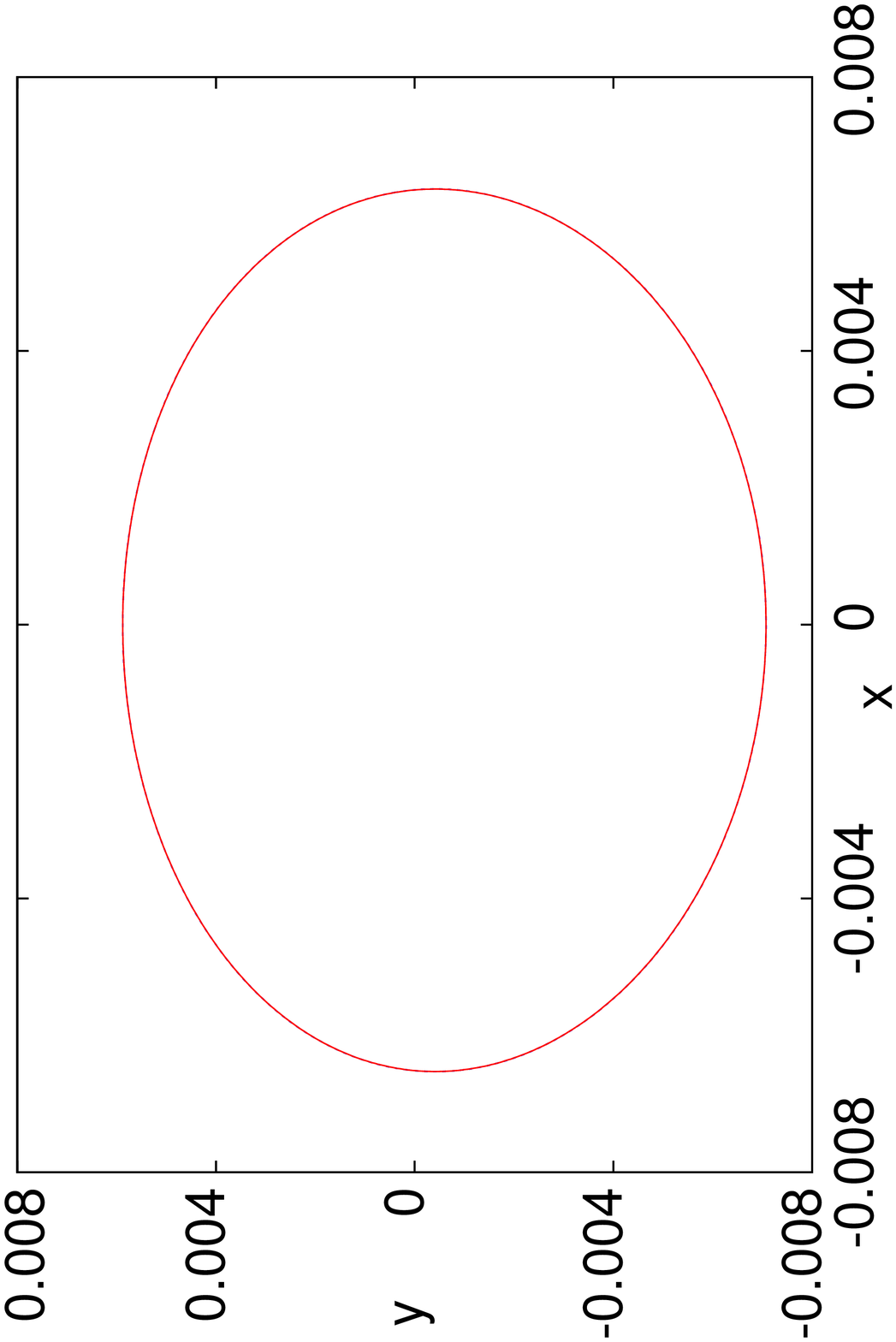}

\put(55.5,34.5){\footnotesize{$\bullet$}}  

\put(28.4,21.5){\color{black}{\vector(1,-1){3}}} 

\put(20.0,57.0){\color{black}(b)}
\end{overpic}

\vspace{ 0.20in} 
\hspace{-0.25in}  
\begin{overpic}[width=0.3\textwidth,height=0.322\textheight,angle=-90]{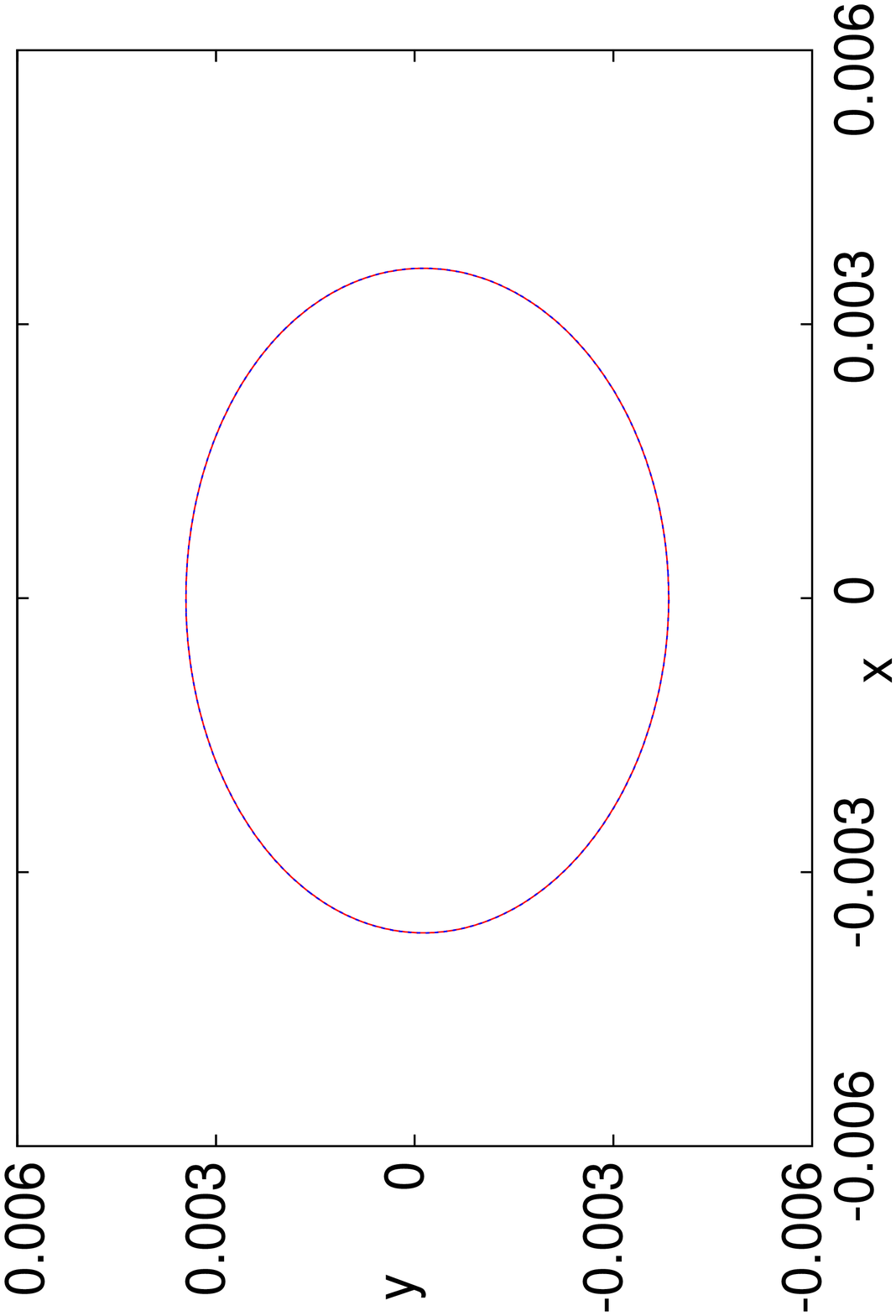}

\put(55.5,34.5){\footnotesize{$\bullet$}}  

\put(35.4,25.8){\color{black}{\vector(1,-1){3}}} 

\put(20.0,57.0){\color{black}(c)}
\end{overpic}
\hspace*{0.47in}
\begin{overpic}[width=0.3\textwidth,height=0.3\textheight,angle=-90]{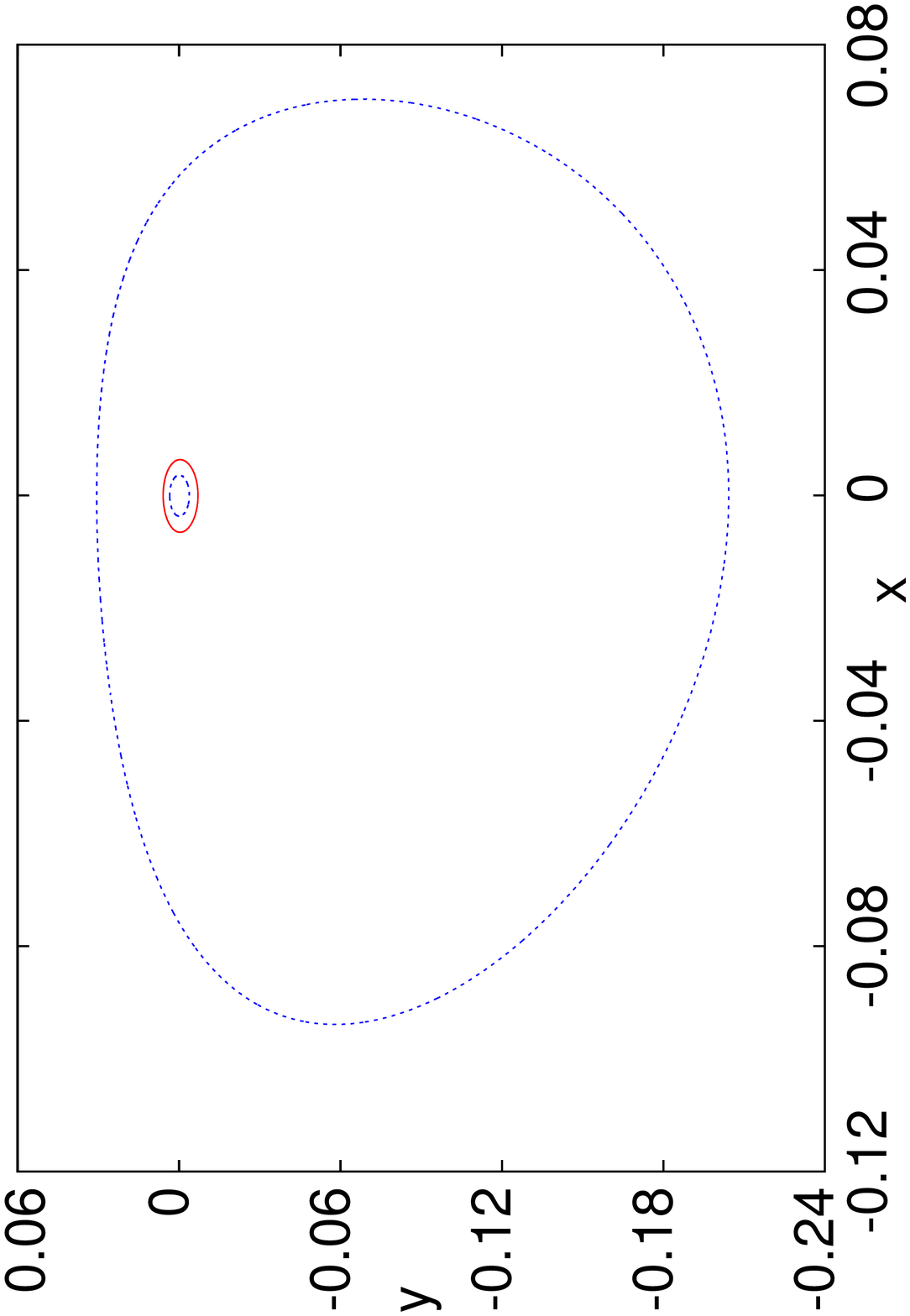}

\put(61.2,56.58){\tiny{$\bullet$}}  

\put(35.4,23.3){\color{black}{\vector(3,-2){5}}} 

\put(15.0,61.0){\color{black}(d)}
\end{overpic}

\vspace{0.10in} 
\caption{Simulation of $3$ small limit cycles using the 
R-K $4$th-order method for system \eqref{b10} 
with $\lambda=-2 \! \times \! 10^{-8}$, 
$ \epsilon =-0.001$ and $\delta =-0.1$: 
(a) the most outer limit cycle with time step $\Delta t = -0.001$, 
and initial points $(0,-3)$ and $(0,0.006)$; 
(b) the middle limit cycle with $\Delta t = 0.01$, 
and initial points $(0,0.006)$ and $(0,0.004)$; 
(c) the most inner limit cycle with $\Delta t = -0.01$, 
and initial points $(0,0.004)$, $(0,0.001)$; and 
(d) three small limit cycles with the stable one in red color and 
unstable ones in blue color.}
\label{fig3}
\end{center}
\end{figure}

To simulate the small limit cycles, the parameter values used by 
Shi, given in \eqref{b11},
to prove the existence of four limit cycles cannot be used 
for simulation since they are too small. When these parameters 
vanish, the focus values are 
$$ 
v_0 = v_1 = v_2 = 0, \quad v_3 = \dfrac{35625}{8}. 
$$
Since $v_3$ is pretty large, we must choose the parameter values such that 
the three limit cycles are very small. 
To achieve this, we choose the following parameter values: 
\begin{equation}\label{b13}
\lambda = -\,\dfrac{2}{10^{8}}, \quad \epsilon = -\,\dfrac{1}{1000}, \quad 
\delta = - \,\dfrac{1}{10}, 
\end{equation} 
which are much larger than that used by Shi given in \eqref{b11}. 
With these parameter values, we apply the Maple program~\cite{Yu1998} 
to obtain the following focus values: 
$$ 
v_0 = - \,\dfrac{1}{10^{8}}, \quad v_1 = \dfrac{1}{10^{3}}, \quad 
v_2 =  -\,\dfrac{2079402109}{112500000}, \quad 
v_3 = \dfrac{59143866813736153313}{16200000000000000}. 
$$ 
Then, the truncated normal form up to $3$rd-order terms is given by 
\begin{equation}\label{b14}  
\dot{r} = v_0 + v_1\, r^2 + v_2\, r^4 + v_4\, r^6. 
\end{equation}
Solving $\dot{r}=0$ yields the approximation of the amplitudes of 
the three small limit cycles as follows: 
$$ 
r_1 \approx 0.003636, \quad r_2 \approx 0.006431, \quad r_3 \approx 0.070769. 
$$ 
The simulation of the three small limit cycles are shown in 
Figure~\ref{fig3}, which are obtained using the R-K $4$th-order method since 
Matlab solver ODE23 (or ODE45) takes too much time to get convergence.  
For a clear view, we only present the data of forming 
the limit cycles, but for each of the limit cycles we 
show two trajectories converging to the same limit cycle, 
one from outside the limit cycle (in blue color) and one from 
inside the limit cycle (in red color).    
It can be seen from Figure~\ref{fig3} that the analytical predictions 
agree very well with the simulations. 

\begin{figure}[!h]
\hspace{0.20in} 
\begin{overpic}[width=0.43\textwidth,height=0.27\textheight,angle=-00]{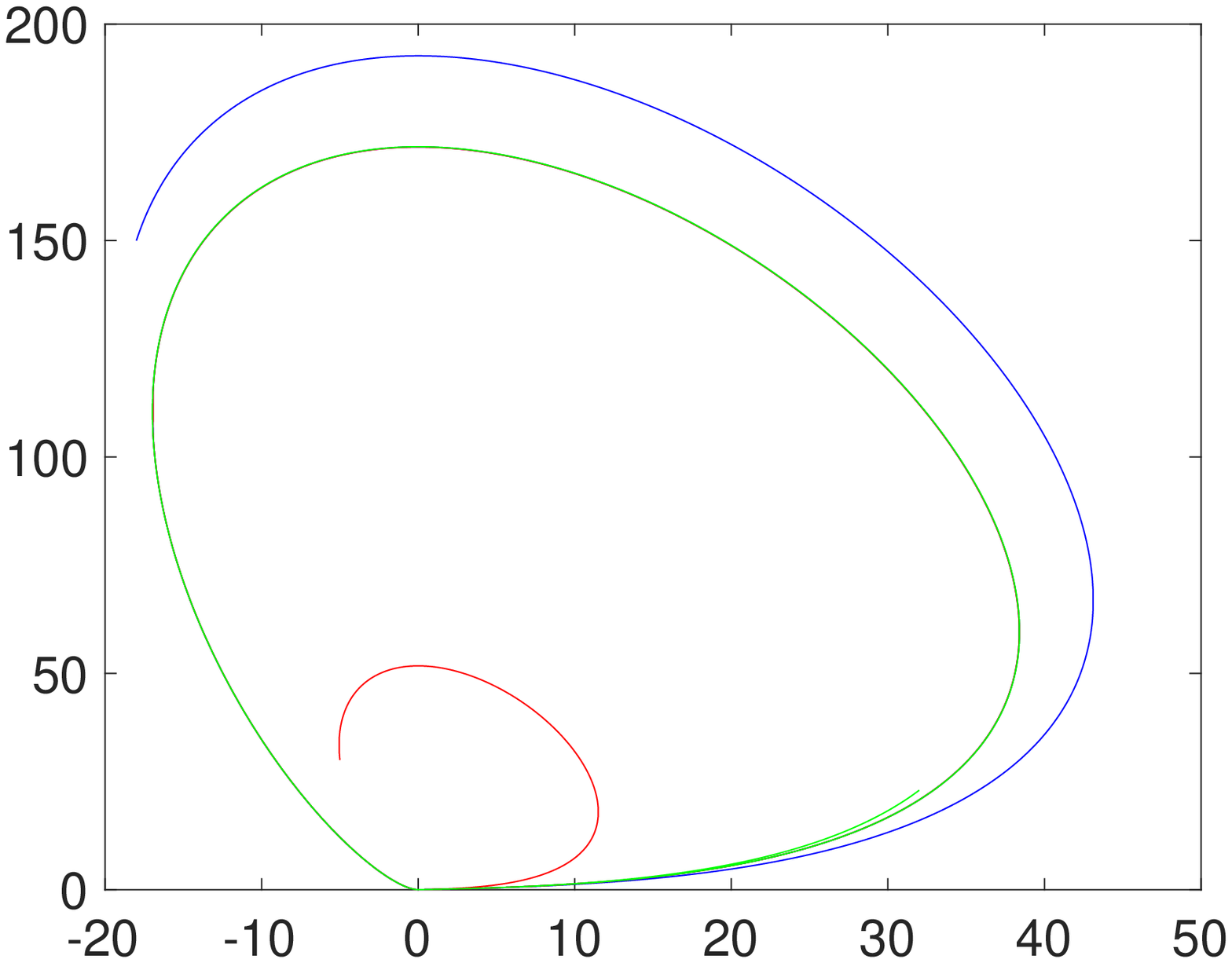} 

\put(14.1,61.4){\color{blue}{\footnotesize{$\bullet$}}}  
\put(28.4,18.8){\color{red}{\footnotesize{$\bullet$}}}  

\put(61.4,67.9){\color{blue}{\vector(1,-1){3}}} 
\put(54.4,64.0){\color{green}{\vector(3,-2){3}}} 
\put(42.4,25.0){\color{red}{\vector(1,-1){3}}} 
 
\put(79.0,70.0){\color{black}(a)}
\end{overpic} 

\vspace*{-2.43in} 
\hspace{3.30in} 
\begin{overpic}[width=0.43\textwidth,height=0.285\textheight,angle=-00]{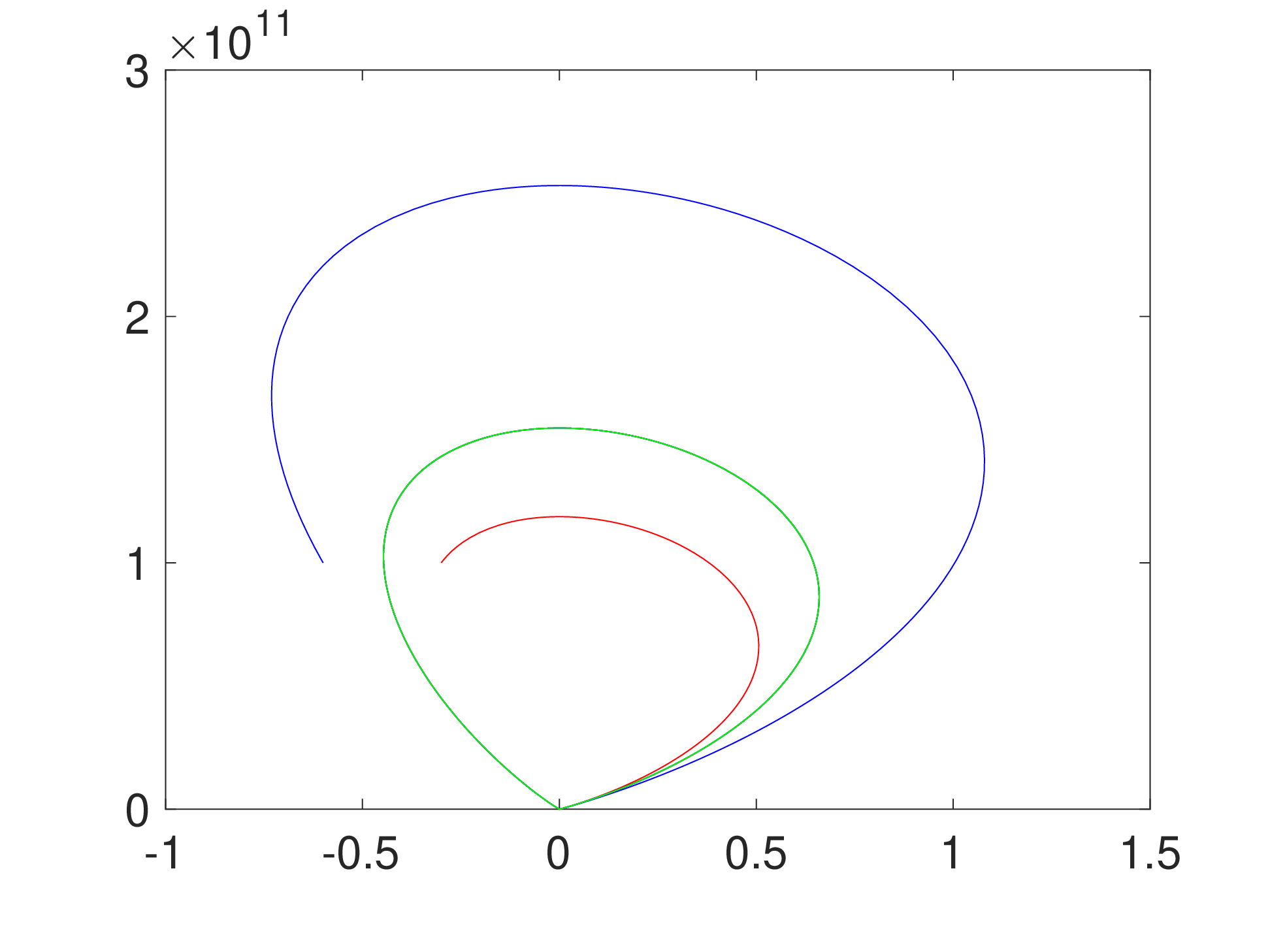}
\put(79.0,74.0){\color{black}(b)}
\put(94.0,7.5){\footnotesize{$\times 10^{11}$} }

\put(24.2,36.0){\color{blue}{\footnotesize{$\bullet$}}}  
\put(33.5,36.0){\color{red}{\footnotesize{$\bullet$}}}  

\put(61.4,69.1){\color{blue}{\vector(2,-1){3}}} 
\put(51.4,49.0){\color{green}{\vector(3,-1){3}}} 
\put(46.4,41.5){\color{red}{\vector(4,-1){3}}} 

\end{overpic}

\vspace{-0.05in}  
\caption{Simulation of the stable big limit cycle using Matlab solver ODE23: 
(a) for system \eqref{b10} taking 
$\lambda=-\,2 \! \times \!  10^{-8}$, $ \epsilon =-\,0.001$, 
$\delta =-\,0.1$, with initial points $(-18,150)$ and 
$(-5,30)$; and (b) for system \eqref{b10} taking 
$\delta_1 = 0.01$, $ \delta_2=0.00002$, with initial points 
$(-0.6 \! \times \! 10^{11},10^{11})$ 
and $(-0.3 \! \times \! 10^{11},10^{11})$.}
\label{fig4}
\end{figure}

With the parameter values given in \eqref{b13}, 
the simulated big limit cycle is obtained using Matlab solver ODE23, 
as shown in Figure~\ref{fig4}(a). Comparing this figure with 
Figure~\ref{fig2}(a) shows that small perturbation parameter  
values can cause obvious change on the size of the limit cycle.

Next, we consider the example proposed by Chen and Wang~\cite{ChenWang}, 
given in \eqref{b12}. 
Setting $\delta_1=\delta_2 =0$ yields the focus values 
$v_0=v_1=0, \ v_2= - \frac{77}{972} \approx -0.079218$.  
Thus we can choose $\delta_1 $ and $\delta_2 $ to obtain two small 
limit cycles with the outer one stable. Then by Poincar\'{e}-Bendixson theory, 
one more small unstable limit cycle can be obtained, which encloses 
the two small limit cycles. For this purpose, we choose 
\begin{equation}\label{b15} 
\delta_1 = 0.01, \quad \delta_2 = 0.00002, 
\end{equation} 
under which 
$$ 
v_0 = -10^{-5}, \quad v_1 = 0.0025, \quad 
v_2 = -\dfrac{172948799}{2332800000} \approx -0.074138. 
$$ 
Then the truncated normal form $ \dot{r}= -10^{-5} + 0.0025\, r^2 
- 0.074138\, r^4 $ gives the approximation of the amplitudes of the 
two small limit cycles as $r_1 = 0.068102$ and $r_2 = 0.170538$. 

With the parameter values, the simulated big limit cycle is obtained 
using Matlab solver ODE23, as shown in Figure~\ref{fig4}(b). 
Comparing this figure with 
Figure~\ref{fig2}(b) again shows that very small perturbation parameter  
values can have great influence on the size of the limit cycle. 
For this case, the limit cycle obtained with 
$\delta_1=0.01,\, \delta_2=0.00002$ 
is only about $1/3$ of that obtained with $\delta_1 = \delta_2 = 0$. 

\begin{figure}[!h]
\begin{center}
\begin{overpic}[width=0.3\textwidth,height=0.3\textheight,angle=-90]{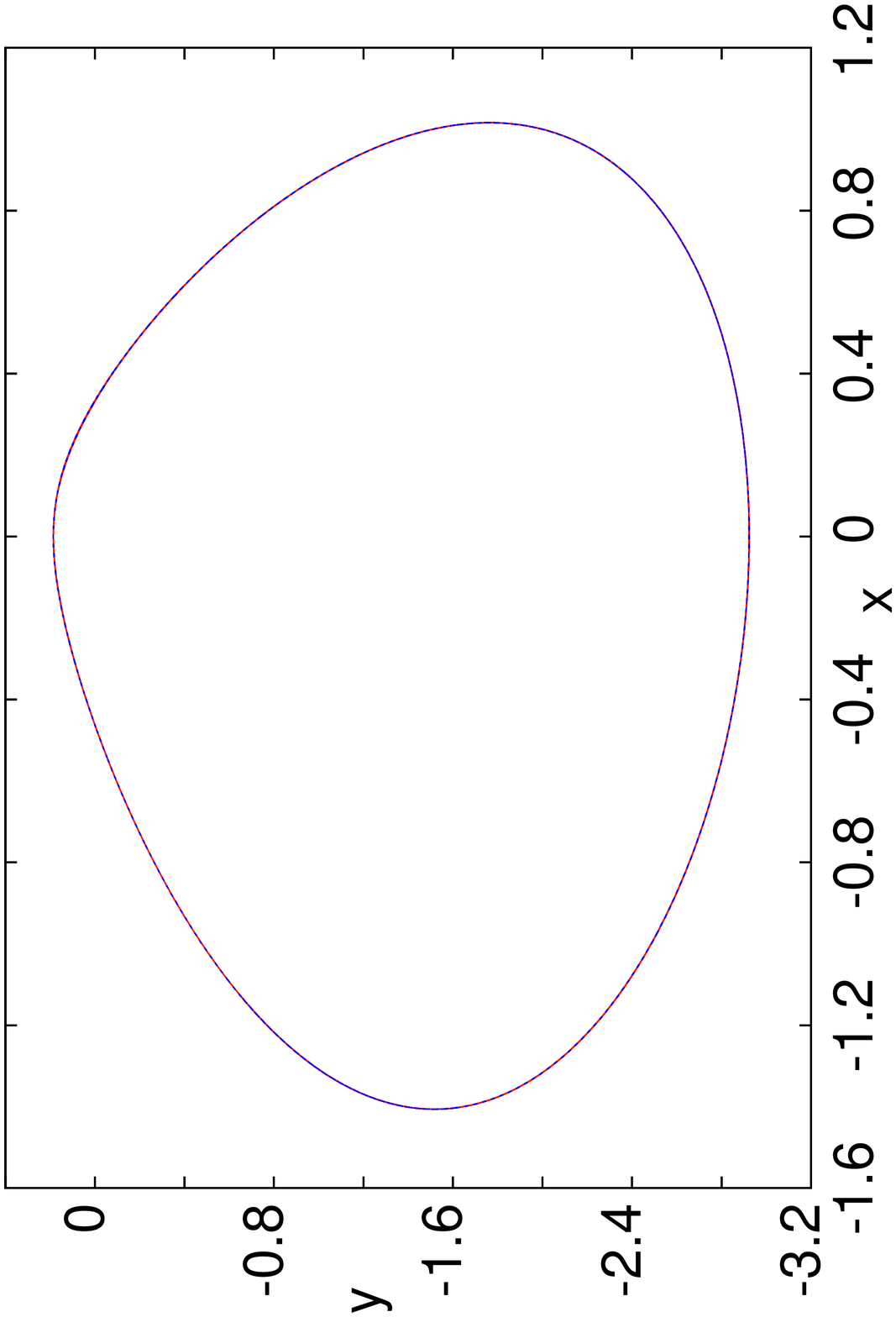}

\put(57.9,61.5){\footnotesize{$\bullet$}}  

\put(28.4,18.8){\color{black}{\vector(2,-1){3}}} 

\put(15.0,62.0){\color{black}(a)}
\end{overpic}
\hspace*{0.50in}
\begin{overpic}[width=0.3\textwidth,height=0.3003\textheight,angle=-90]{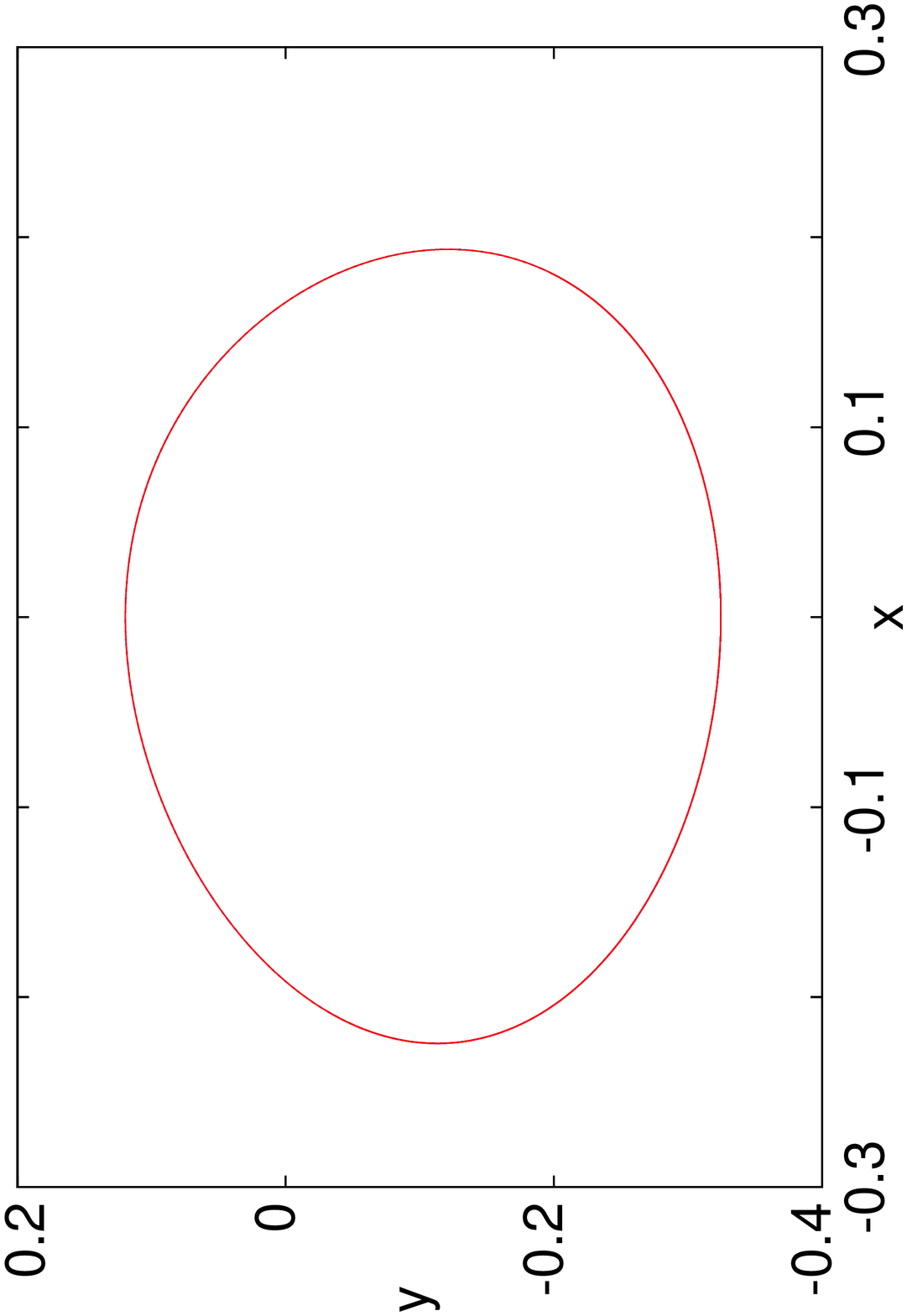}

\put(51.8,47.7){\footnotesize{$\bullet$}}  

\put(28.4,22.2){\color{black}{\vector(3,-2){3}}} 

\put(16.0,61.0){\color{black}(b)}
\end{overpic}

\vspace{ 0.20in} 
\hspace{-0.09in}  
\begin{overpic}[width=0.3\textwidth,height=0.303\textheight,angle=-90]{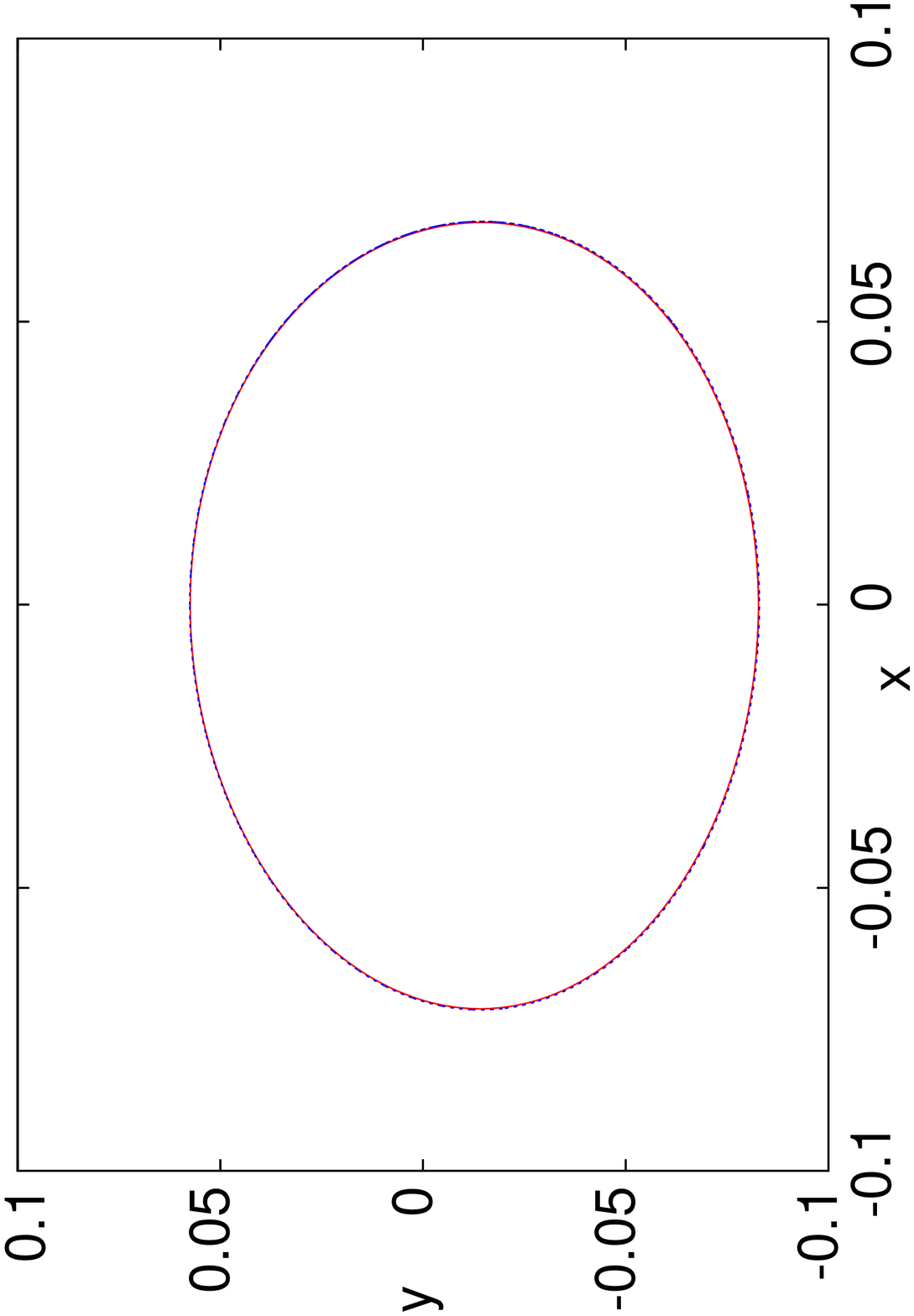}

\put(52.3,36.7){\footnotesize{$\bullet$}}  

\put(30.4,19.2){\color{black}{\vector(3,-2){3}}} 

\put(16.0,61.0){\color{black}(c)}
\end{overpic}
\hspace*{0.50in}
\begin{overpic}[width=0.3\textwidth,height=0.3\textheight,angle=-90]{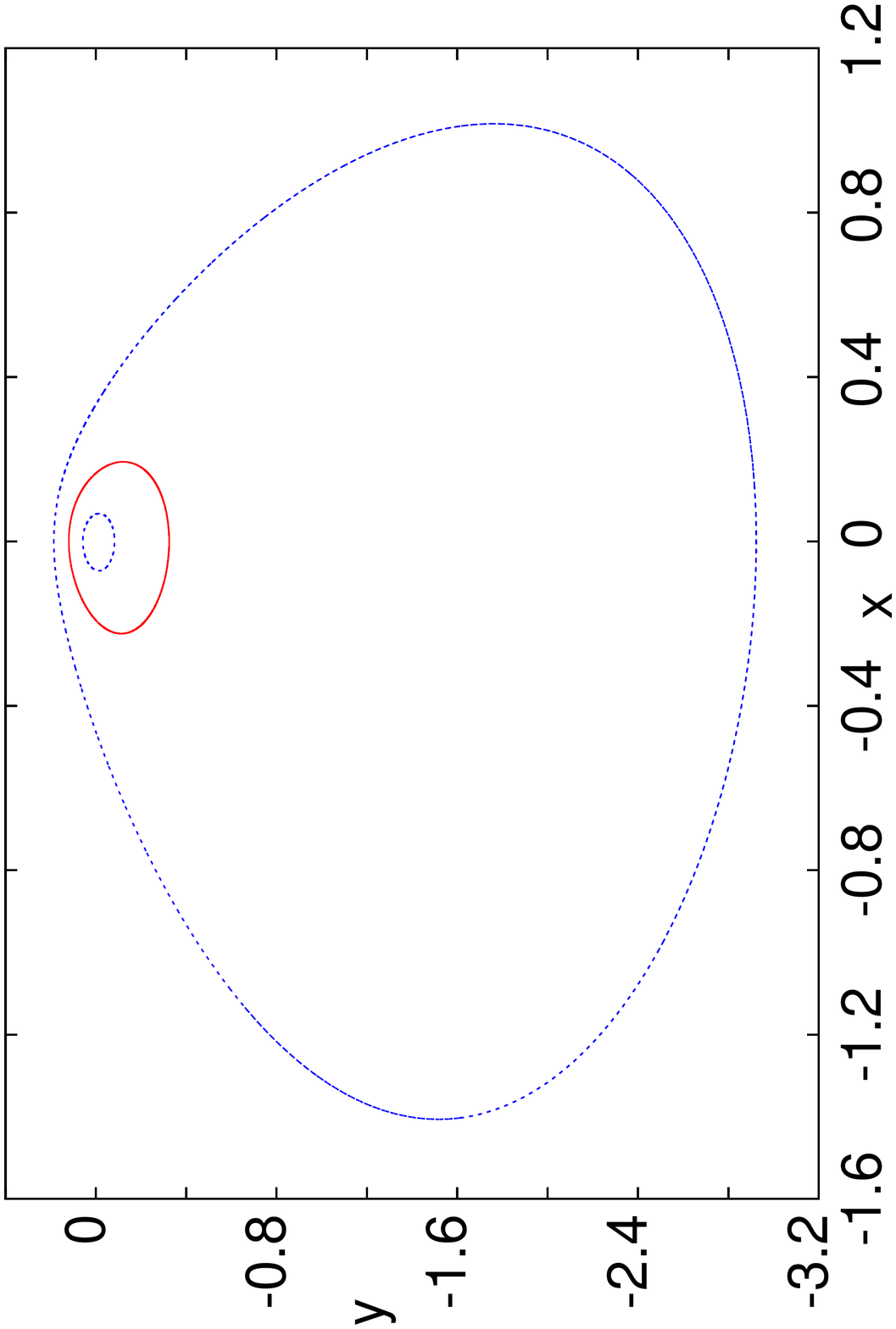}

\put(58.18,62.1){\tiny{$\bullet$}}  

\put(30.4,17.8){\color{black}{\vector(2,-1){3}}} 
\put(58.3,57.4){\color{black}{\vector(1,0){2}}} 

\put(15.0,61.0){\color{black}(d)}
\end{overpic}

\vspace{0.10in} 
\caption{Simulation of three small limit cycles using the 
R-K $4$th-order method for system \eqref{b12} 
with $ \delta_1 =0.01$ and $\delta_2 =0.00002$: 
(a) the most outer limit cycle with time step $\Delta t = -0.001$, 
and initial points $(0,0.22)$ and $(0,0.15)$; 
(b) the middle limit cycle with $\Delta t = 0.001$, 
and initial points $(0,0.15)$ and $(0,0.06)$; 
(c) the most inner limit cycle with $\Delta t = -0.001$, 
and initial points $(0,0.06)$, $(0,0.01)$; and 
(d) three small limit cycles with the stable one in red color and 
unstable ones in blue color.}
\label{fig5}
\end{center}
\end{figure}

The simulated three small limit cycles are obtained using the 
R-K $4$th-order method, as shown in 
Figure~\ref{fig5}. Again for a clear view, for each of the limit cycles 
we show two trajectories converging to the same limit cycle, 
one from outside the limit cycle (in blue color) and one from 
inside the limit cycle (in red color).    
It can be seen from this figure that the analytical predictions 
for the two smaller limit cycles agree very well with the simulations.

\section{Four big size limit cycles obtained in~\cite{KKL2013}}

\begin{figure}[!h]
\hspace{0.20in} 
\begin{overpic}[width=0.43\textwidth,height=0.27\textheight,angle=-00]{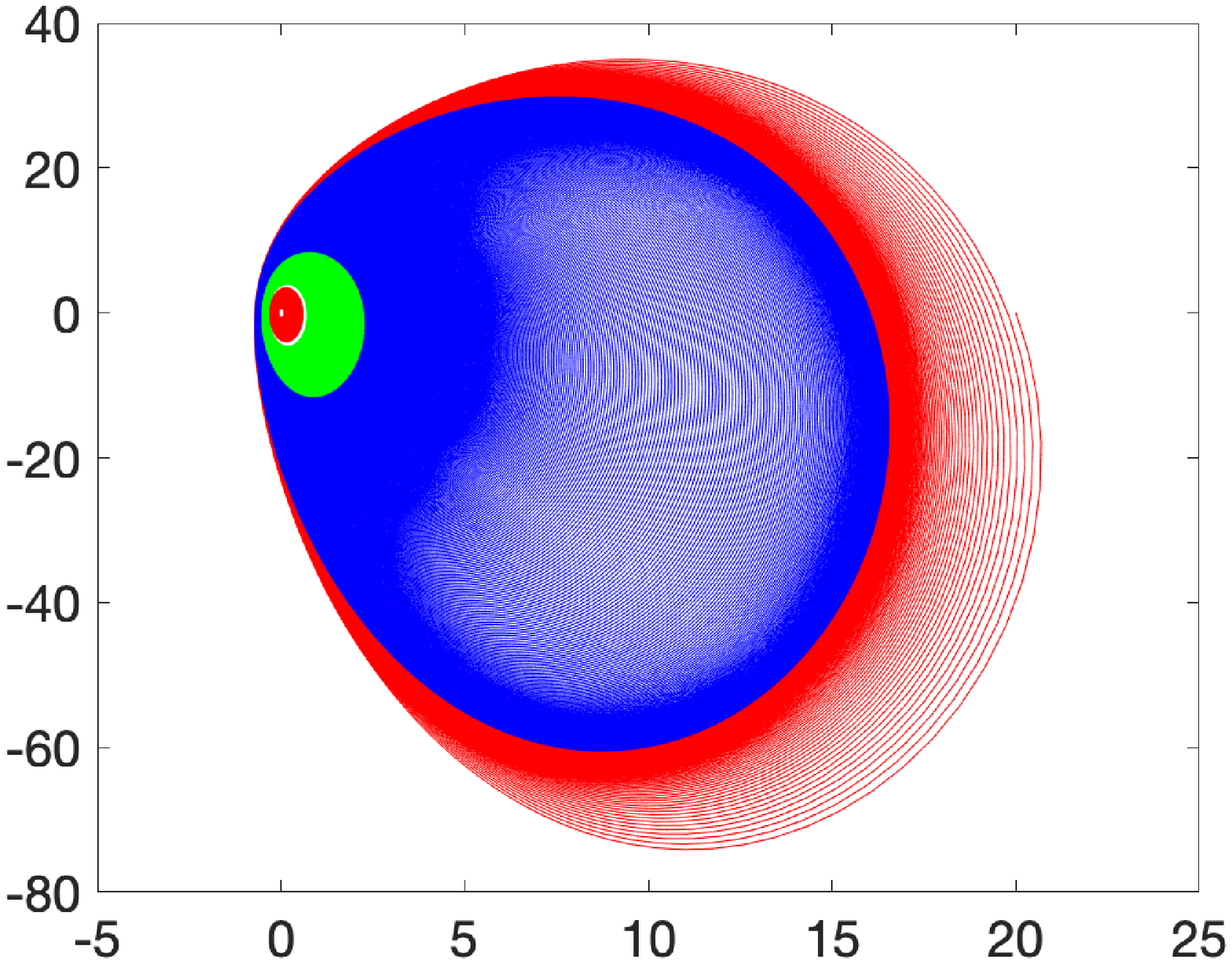} 
\put(79.0,70.0){\color{black}(a)}
\end{overpic} 

\vspace*{-2.365in} 
\hspace{3.30in} 
\begin{overpic}[width=0.43\textwidth,height=0.27\textheight,angle=-00]{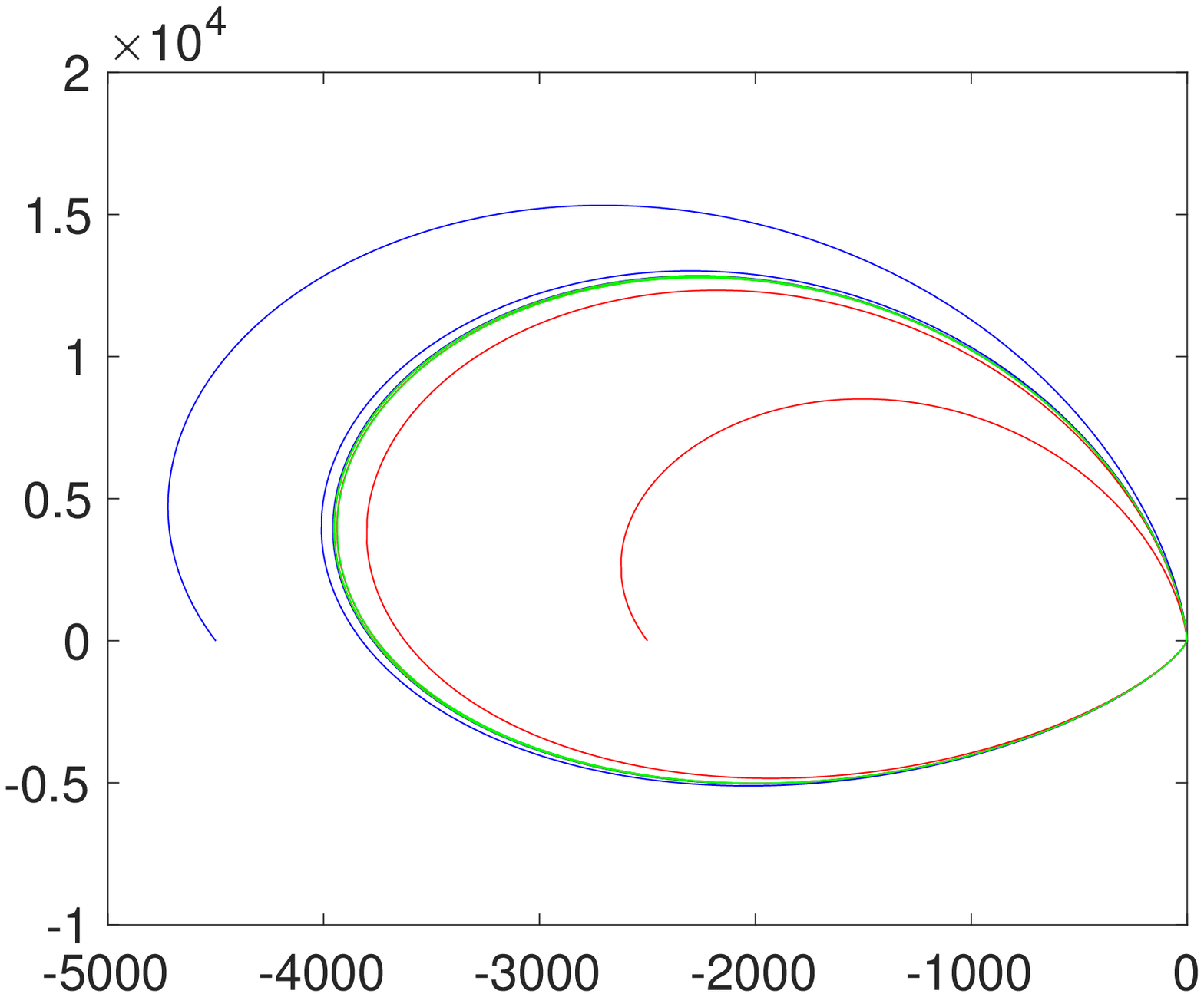}
\put(79.0,70.0){\color{black}(b)}

\put(19.3,32.0){\color{blue}{\footnotesize{$\bullet$}}}  
\put(50.2,32.0){\color{red}{\footnotesize{$\bullet$}}}  

\put(48.4,69.0){\color{blue}{\vector(-1,0){2}}} 
\put(35.0,55.0){\color{green}{\vector(-1,-1){2}}} 
\put(67.3,52.9){\color{red}{\vector(-1,0){2}}} 
\end{overpic}

\caption{Simulation of limit cycles using Matlab solver ODE23 
for system \eqref{b8} with the parameter values given in \eqref{b17}: 
(a) three limit cycles around the unstable focus 
$(0,0)$; and (b) one big unstable limit cycle (in green color)  
around the stable focus $(-6.2596, \, 7.4498)$, with 
two trajectories starting from the initial points 
$(-4500,0)$ (in blue color) and $(-2500,0)$ (in red color).} 
\label{fig6}
\end{figure}

In this section, we consider system \eqref{b8}. 
In \cite{KKL2013}, Kuznetsov {\it et al.} proved that the system has 
four limit cycles if the following conditions hold: 
\begin{equation}\label{b16} 
\begin{array}{ll}  
b_2 \in (1,3),  \quad 
c_2 \in \Big( \dfrac{1}{3},\,1 \Big), \quad 
4 a_2 (c_2-1) > (b_2-1)^2, \quad b_2 c_2 >1, \\[2.0ex] 
\alpha_2 \in \Big( \dfrac{a_2(b_2+2)}{b_2 c_2 -1}, \ 
 \dfrac{a_2(b_2+2)}{b_2 c_2 -1} + \delta \Big), \quad 
\beta_2 \in (0, \varepsilon), \quad 
0 < \varepsilon \ll \delta \ll 1. 
\end{array} 
\end{equation}   
The following parameter values:
\begin{equation}\label{b17} 
a_2 =-10, \quad b_2 = 2.2, \quad c_2 = 0.7, \quad 
\alpha_2 = -72.7778, \quad \beta=0.0015, 
\end{equation} 
were chosen in \cite{KKL2013} to obtain four big size limit cycles. 
With the above parameter values, system \eqref{b8} has 
two fixed points, 
$$ 
{\rm E_0}=(0,0) \quad \textrm{and} \quad 
{\rm E_1}= (-6.2596, \, 7.4498). 
$$ 
We use the parameter values in \eqref{b17} and apply Matlab solver ODE23 to 
obtain the simulated three limit cycles around ${\rm E_0}$ as shown 
in Figure~\ref{fig6}(a),  
and one big unstable limit cycle around ${\rm E_1}$ 
as shown in Figure~\ref{fig6}(b). Note that the three limit cycles 
shown in Figure~\ref{fig6}(a) are those loops between different 
colors. The outside red trajectory converges to the most outer limit cycle, 
while the blue trajectory ``converges'' to the middle unstable limit cycle 
using backward time integration, the green trajectory converges 
to the most inner stable limit cycle, and the inside red trajectory 
``converges'' to the unstable focus ${\rm E_0}$ using backward time 
integration. In Figure~\ref{fig6}(b), with backward time integration, 
two trajectories ``converge'' to the unstable big limit cycle 
(in green color), one from outside (in blue color) and one from 
inside (in red color).   

%

%
%
%
%
%
%

\section{Four normal size limit cycles obtained for the 
near-integrable system~\eqref{b9}}

Finally, we study the bifurcation of four limit cycles in 
near-integrable system \eqref{b9}. This system has two 
centers at $(0,0)$ and $(0,1)$ at $\varepsilon=0$.  
The Melnikov function method has been used in \cite{YuHan2012} to 
show that when $a_1 < -\,1$, system (\ref{b9}) can have small limit 
cycles bifurcating from the two centers $(0,0)$ and $(1,0)$ with 
distributions: $(3,0)$, $(0,3)$, $(2,0)$, $(0,2)$ and $(1,1)$. 
The Melnikov functions associated with the two centers are respectively 
given by 
\be
\begin{array}{rl}  
M_0(h, a_{ij}, b_{ij}) = & \mu_{00} \, ( h- h_{00}) 
+ \mu_{01} \, (h-h_{00})^2 
+ \mu_{02} \, (h-h_{00})^3 
\\ [1.0ex]  
& +\, \mu_{03} \, (h-h_{00})^4 
+ O((h-h_{00})^5), \quad  {\rm for} \ \ 0 < h - h_{00} \ll 1, 
\\[1.5ex] 
M_1(h,  a_{ij}, b_{ij}) = & \mu_{10} \, ( h_{10} -h) 
+ \mu_{11} \, (h_{10}-h)^2 
+ \mu_{12} \, (h_{10}-h)^3 
\\ [1.0ex]  
& +\, \mu_{13} \, (h_{10}-h)^4 
+ O((h_{10}-h)^5), \quad  {\rm for} \ \ 0 < h_{10}-h \ll 1, 
\end{array} 
\l{b18}
\ee
where
\be 
\begin{array}{ll} 
L_h: \ H(x,y) = h \left\{ 
\begin{array}{ll}  
\in (h_{00}, \infty), & {\rm for} \ \ 1+a_1\,x > 0, \\[0.5ex]  
\in (-\infty,h_{10}), & {\rm for} \ \ 1+a_1\,x < 0,  
\end{array}  
\right.  
\\[2.5ex]  
h_{00} = 
H(0,0) = \dss\frac{1+a_1-a_4}{2\, a_4\, (a_1-a_4)\,(a_1 - 2 a_4)},  
& {\rm for} \ \ 1+a_1\,x > 0, \\[2.0ex] 
h_{10} = H(1,0) = -\, \dss\frac{(a_1+1)\,(a_4+1)}
{2\, a_4\, (a_1-a_4)\,(a_1 - 2 a_4)}\, (-1-a_1)^{- \frac{2 a_4}{a_1}}, 
\quad & {\rm for} \ \ 1 + a_1\,x < 0, 
\end{array}
\l{b19} 
\ee   
and the coefficients $\, \mu_{ij}, \ i=0,\,1; \ j=0,\,1,\,2,\, \cdots \,$ 
can be obtained by using the Maple programs developed 
in~\cite{HanYangYu2009}. It should be noted that 
all the coefficients $\mu_{1k},\, k=1,2, \dots $ obtained in 
\cite{HanYangYu2009} should be multiplied by $-1$, since 
the rotations along the loop $H(x,y)=h$ on the right side and 
left side of the line $x=-\frac{1}{a_1}$ have opposite direction. 
But this error does not affect the conclusion on the number of 
bifurcating limit cycles.  

Further, the following results were proved in \cite{YuHan2012}. 
For the case of bifurcation of small limit cycles from the two centers 
$(0,0)$ and $(1,0)$ with $(3,0)$-distribution 
(respectively, $(0,3)$-distribution) there exists at least one 
large limit cycle near $L_h$ for some 
$\, h \in (- \infty, h_{10})$ (respectively for 
some $\, h \in (h_{00}, \infty)$). 
For the case of limit cycles with $(2,0)$-distribution 
(respectively, $(0,2)$-distribution) there exist at least two 
large limit cycles, one near $L_{h_1}$ for some $\, h_1 \in (- \infty, h_{10})$ 
and one near $L_{h_2}$ for some $\, h_2 \in (h_{00}, \infty)$.   
There exist corresponding values of the parameters $\,a_1 \,$ and $\, a_4\,$ 
for the existence of four limit cycles.  

Four cases were considered in \cite{YuHan2012} to yield four limit 
cycles, described as below. 

\begin{enumerate} 
\item[{(A)}] 
$(3,0)$-distribution for small limit cycles plus a 
$(0,1)$-distribution of large limit cycle, resulting in 
$(3,1)$-distribution, with the parameter values given by 
$$  
\begin{array}{ll} 
a_1 = -\dfrac{30}{7}, \quad 
a_4 =  \dfrac{1}{3}\, (a_1-5) - \varepsilon_1, \\[2.0ex] 
b_{11} =  \dfrac{(a_1+2 a_4)(1+a_4-a_1)}{1+a_4} \, a_{10} - \varepsilon_2, 
\quad b_{01} = -a_{10} - \varepsilon_3 , 
\end{array}
$$  
where $\varepsilon_k,\,k=1,2,3$ are perturbation parameters. 

\item[{(B)}]
$(0,3)$-distribution for small limit cycles plus 
a $(1,0)$-distribution of large limit cycle, resulting 
in a $(1,3)$-distribution, with the parameter values given by 
$$
\begin{array}{ll} 
a_1 = -\dfrac{70}{51}, \quad 
a_4 =  \dfrac{1}{3}\, (6a_1+5) - \varepsilon_1, \\[2.0ex]  
b_{11} =  \dfrac{(a_1+2 a_4)(2 a_1 -a_4+1)}{(1+a_1)^2(a_1-a_4+1)} \, a_{10} 
- \varepsilon_2, 
\quad b_{01} = -b_{11} + \dfrac{2 a_4-1}{1+a_1}\, a_{10} - \varepsilon_3.  
\end{array}
$$ 

\item[{(C)}]
$(2,0)$-distribution for small limit cycles 
plus a $(1,1)$-distribution of large limit cycles, 
resulting in a $(3,1)$-distribution, with the parameter values given by 
$$ 
a_1 = -4, \quad a_4 = - \dfrac{18}{5} - \varepsilon_1, \quad 
b_{11} = \dfrac{392}{65}\, a_{10} - \varepsilon_2, \quad 
b_{01} = - a_{10} - \varepsilon_3. 
$$

\item[{(D)}]
$(0,2)$-distribution for small limit cycles plus 
a $(1,1)$-distribution of large limit cycles, resulting in 
$(1,3)$-distribution, with the parameter values given by 
$$
a_1 = -\, \dfrac{4}{3}, \quad a_4 = -\, \dfrac{6}{5} - \varepsilon_1,  
\quad b_{11} = \dfrac{1176}{65}\, a_{10} - \varepsilon_2, \quad 
b_{01} = -\, \frac{513}{65} \, a_{10} - \varepsilon_3. 
$$ 
\end{enumerate} 
It should be pointed out that although the above formulas 
were given in \cite{YuHan2012} and the existence of large limit cycles 
were proved using the Melnikov function for each case, the 
three small limit cycles were not explicitly shown but with a proof 
for the existence of fine focus with a schematic plotting. In other words, 
the three perturbations $\varepsilon_k, \, k=1,2,3$ were not definitely 
defined to numerically demonstrate the three small limit cycles.  

Now, we choose the formulas and parameter values given in (A) for simulation. 
When $\varepsilon_k=0,\, k=1,2,3$, the focus values  
associated with $(0,0)$ are 
$$ 
v_0 = v_1 = v_2 = 0, \quad v_3 = \dfrac{347875}{7260624} 
\approx 0.047913.  
$$
Now we set  
$$ 
a_{10} = 0.005, \quad 
\varepsilon_1 = 0.1, \quad \varepsilon_2 = 3 \! \times \! 10^{-5}, \quad 
\varepsilon_3 = 10^{-8}, 
$$ 
which results in 
$$ 
a_1 = -\dfrac{30}{7}, \quad 
a_4 = -\dfrac{671}{210}, \quad 
a_{10} = \dfrac{1}{200}, \quad 
b_{01} = -\dfrac{500001}{100000000}, \quad 
b_{11} = \dfrac{49182857}{968100000}, \quad 
$$
and further taking $\varepsilon=\frac{1}{100}$ 
we have system \eqref{b12} in the form of 
\be  
\begin{array}{rl} 
\dot{x} \!\!\! & = y \, \Big(1 - \dfrac{30}{7}\, x\Big) 
+ \dfrac{1}{100} \times \dfrac{1}{200} \, x, \\ [2.0ex]  
\dot{y} \!\!\! & = -\, x + x^2 -\dfrac{671}{210} \, y^2 
+ \dfrac{1}{100} \times \Big( -\dfrac{500001}{100000000} \, y 
+ \dfrac{49182857}{968100000} \, x\, y \Big). 
\end{array}
\l{b20} 
\ee 
Then, we apply the Maple program \cite{Yu1998} to obtain the 
following focus values: 
$$ 
v_0 = -\frac{1}{200000000}, \ \ 
v_1 = \frac{461}{56000000}, \ \ 
v_2 = -\frac{36481804571}{14817600000000}, \  
v_3 =  \frac{11326448548182069181}{25092716544000000000}. 
$$ 
Then solving the truncated normal form \eqref{b14} with the 
above focus values we obtain the approximated amplitudes of 
the three small limit cycles around $(0,0)$: 
$$
r_1 \approx 0.026385, \quad r_2 \approx 0.079134, \quad 
r_3 \approx 0.140966. 
$$

\begin{figure}[!h]
\vspace{0.00in}
\begin{center}
\begin{overpic}[width=0.28\textwidth,height=0.28\textheight,angle=-90]{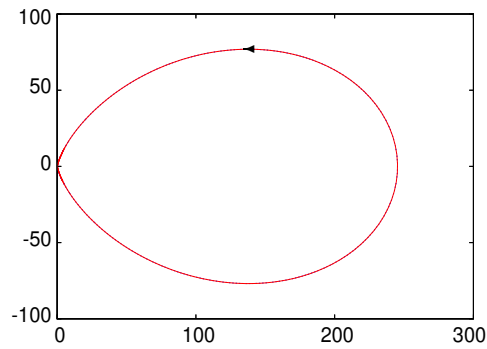}

\put(50.4,62.9){\color{black}{\vector(-1,0){2}}} 

\end{overpic}
\caption{Simulated two trajectories for system \eqref{b20},
converging to the big stable limit cycle around the unstable focus $(1,0)$,
with the initial points $(300,0)$ (in red color) and 
$(10,0)$ (in blue color).}
\label{fig7}
\end{center}
\end{figure}

\begin{figure}[!h]
\begin{center}
\begin{overpic}[width=0.28\textwidth,height=0.28\textheight,angle=-90]{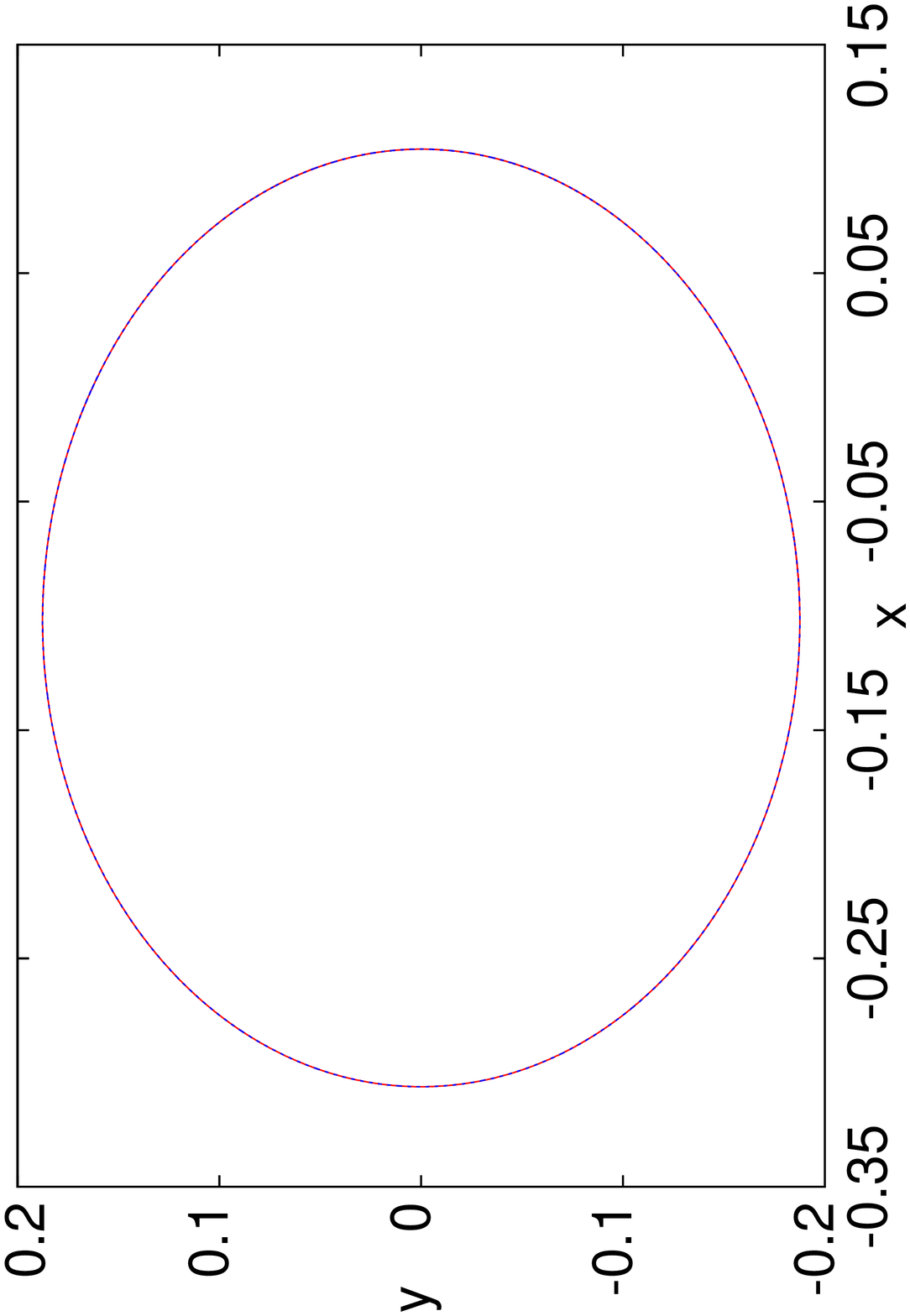}

\put(69.1,37.0){\footnotesize{$\bullet$}}  

\put(77.4,60.1){\color{black}{\vector(1,-1){2}}} 

\put(13,60){(a)} 
\end{overpic} 
\hspace{0.30in} 
\begin{overpic}[width=0.28\textwidth,height=0.283\textheight,angle=-90]{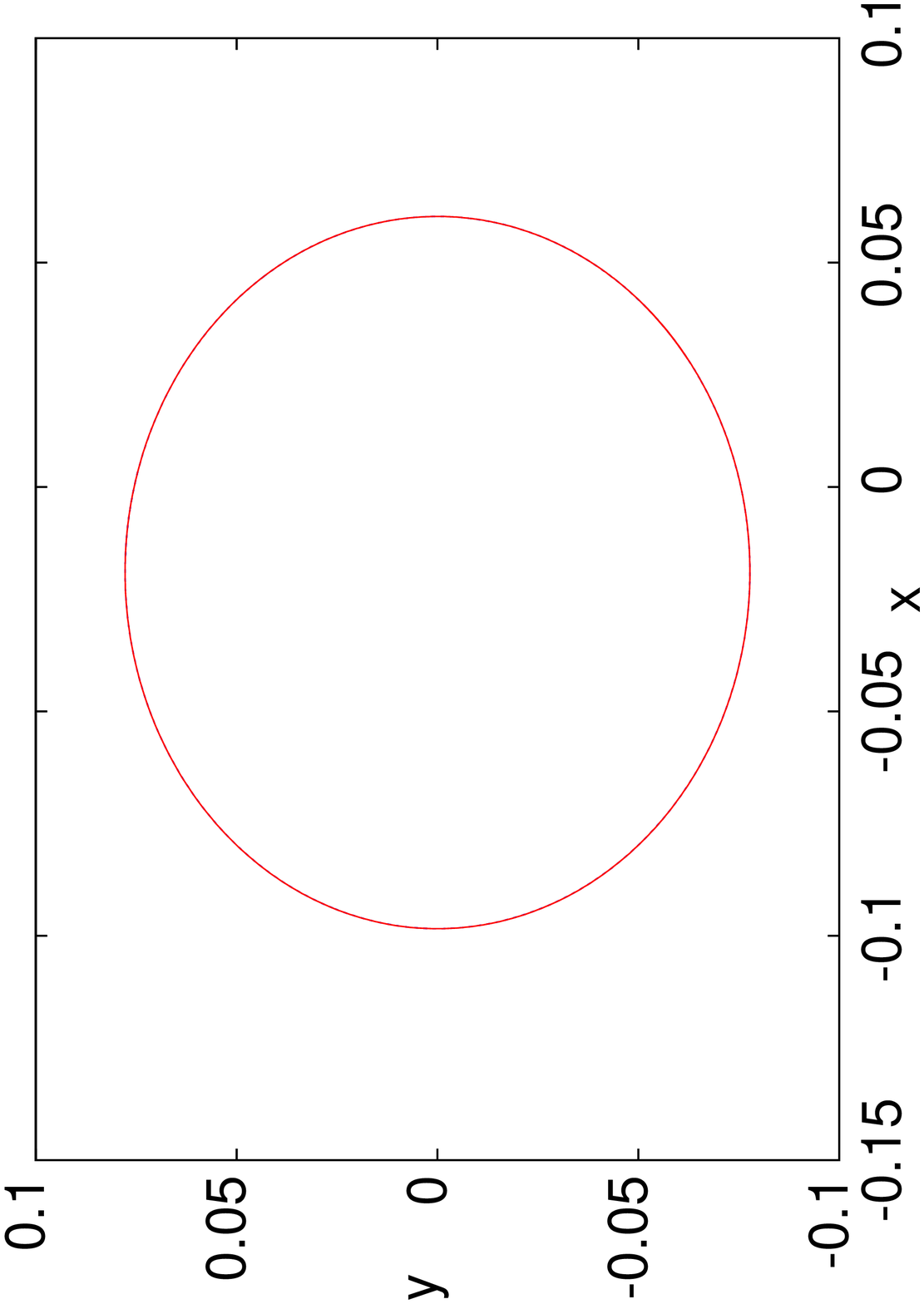}

\put(61.0,36.7){\footnotesize{$\bullet$}}  

\put(73.7,56.4){\color{black}{\vector(1,-1){2}}} 

\put(14,60){(b)} 
\end{overpic}

\vspace{0.20in} \hspace{-0.07in} 
\begin{overpic}[width=0.28\textwidth,height=0.283\textheight,angle=-90]{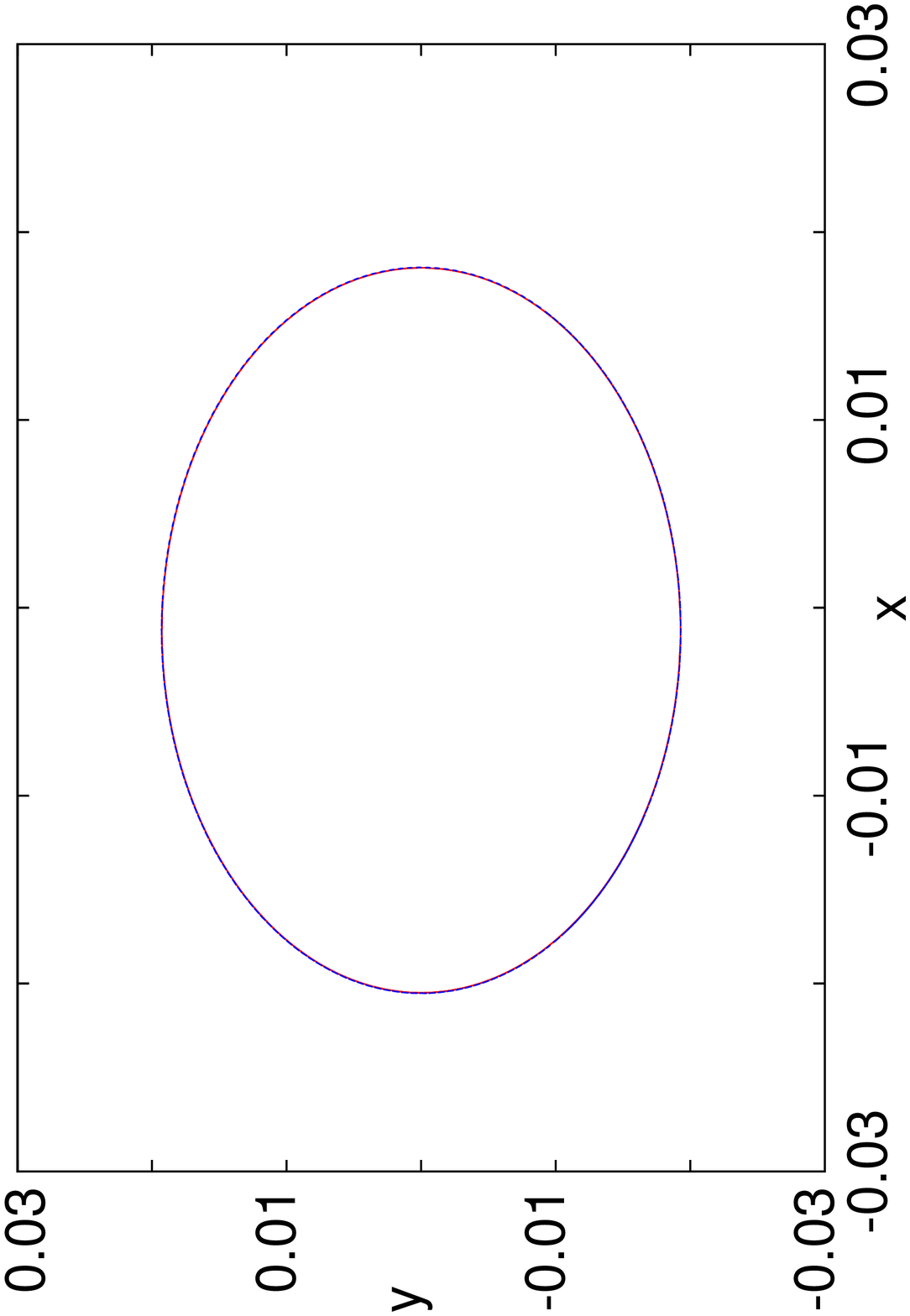}

\put(52.2,36.7){\footnotesize{$\bullet$}}  

\put(70.0,53.2){\color{black}{\vector(3,-2){2}}} 

\put(14,60){(c)} 
\end{overpic}
\hspace{0.34in} 
\begin{overpic}[width=0.28\textwidth,height=0.28\textheight,angle=-90]{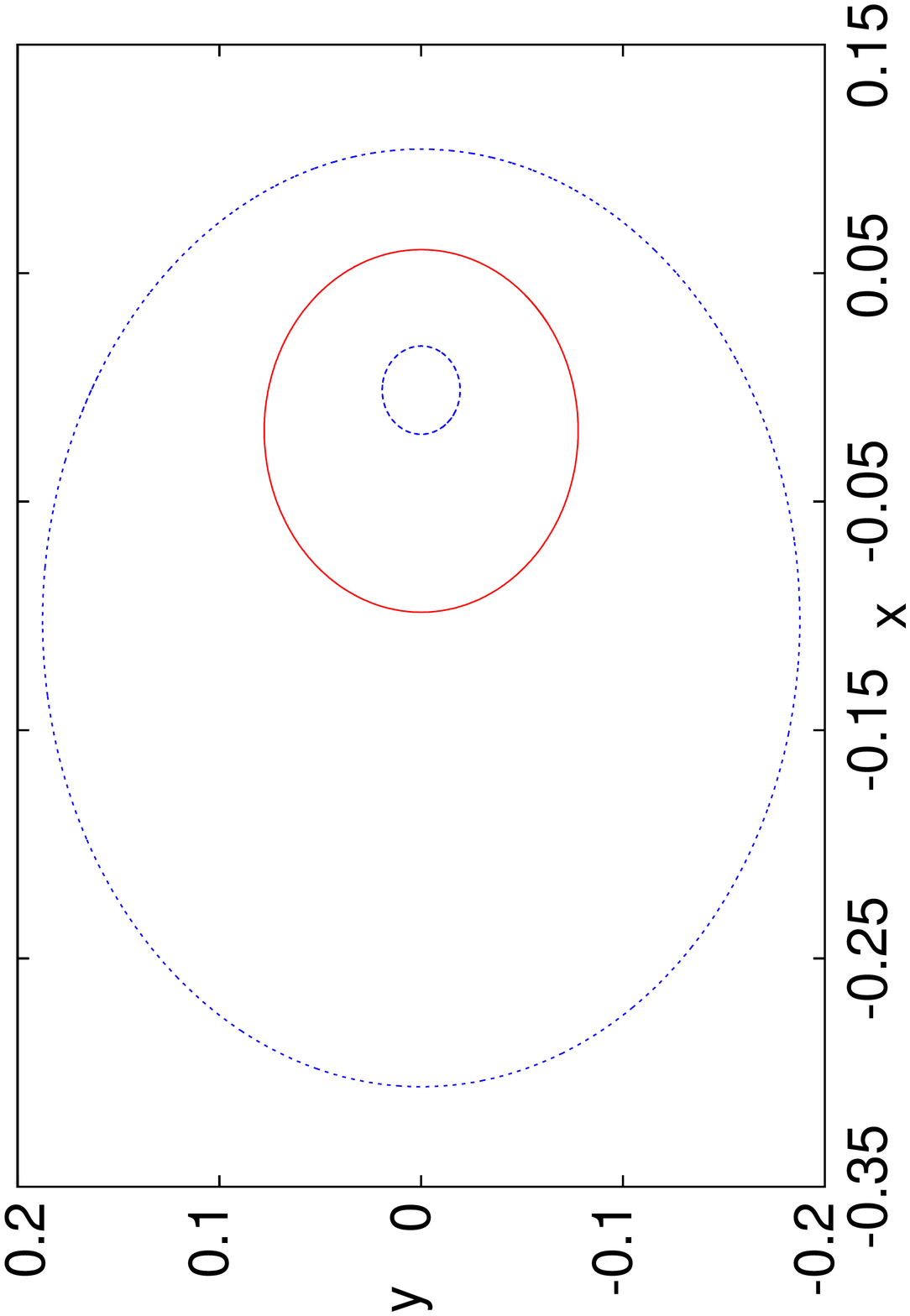}

\put(69.1,36.75){\footnotesize{$\bullet$}}  

\put(73.0,63.0){\color{black}{\vector(3,-2){2}}} 
\put(71.0,50.2){\color{black}{\vector(3,-1){2}}} 
\put(69.8,41.2){\color{black}{\vector(1, 0){2}}} 

\put(13,60){(d)} 
\end{overpic}

\vspace{0.10in} 
\caption{Simulation of the three small limit cycles for system \eqref{b20}: 
(a) two trajectories ``converging'' (backward time) to the most outer 
unstable limit cycle, from the initial points $(-0.35,0)$ (in red color)
and $(-0.25,0)$ (in blue color); 
(b) two trajectories converging to the middle 
stable limit cycle, from the initial points $(-0.25,0)$ (in red color)
and $(-0.05,0)$ (in blue color); 
(c) two trajectories ``converging'' (backward time) to the most inner 
unstable limit cycle, from the initial points $(-0.05,0)$ (in red color)
and $(-0.002,0)$ (in blue color); and (d) three limit cycles with stable 
one in red color and unstable ones in blue color.}
\label{fig8}
\end{center}
\end{figure}

However, unlike the three quadratic systems considered in the previous 
two sections, the simulation for this near-integrable system 
is extremely difficult since the convergence speed is too slow. 
We apply the R-K $4$th-order method to simulate the system and 
obtain the results shown in Figures~\ref{fig7} and \ref{fig8}. 
Again, for each of the limit cycles we take two initial points, one from 
outside and one from inside the limit cycle. Also, we use 
the backward time in integration scheme for unstable limit cycles. 
The time step, the number of iterations and the CPU time are given 
in Table 1.

\begin{table}[!h]\label{table1} 
\vspace{-0.00in}
\centerline{Table 1. Simulation data for system \eqref{b20}.}
\vspace{0.10in}
\centering
{\footnotesize
\begin{tabular}{|c|c|c|c|c|}
\hline
  & & & & \vspace{-0.10in} \\
$\begin{array}{c}
\textrm{Trajectories}\\
\textrm{moving towards} 
\end{array}$  & $\begin{array}{c} 
\textrm{Initial} \\
\textrm{point} 
\end{array} $ 
 &  $\begin{array}{c} 
\textrm{Time} \\
\textrm{step} 
\end{array} $  
& $\begin{array}{c} \textrm{Number of} \\ 
\textrm{interation} \end{array} $ & 
$\begin{array}{c} \textrm{CPU} \\ \textrm{time} \end{array} $ \\ 
 &  &  & & \vspace{-0.12in} \\
\hline \hline
 & &  &  &  \vspace{-0.10in} \\
$\infty$ 
& &  $\!-0.0001$ & $10^8$ & 2 sec \\ 
  &  $\begin{array}{c}  
(300,0) 
\end{array} $
 & &  &  \vspace{-0.12in} \\
\cline{1-1}\cline{3-5} 
  & & &  & \vspace{-0.10in} \\
\hspace{-0.10in} 
$\begin{array}{c} 
\textrm{The large LC} \\ 
\textrm{(Stable)} \\[-3.0ex]  
\end{array} $ 
 & & $\begin{array}{c} 
\ \ 0.0001 \\[-3.0ex]  
\end{array}$ 
& 
$\begin{array}{c} 
\\[-0.0ex] 
10^{12} \\[0.0ex]  
\end{array}$  & 
$\begin{array}{c} 
\\[-0.0ex]  
12 \ \textrm{hr} \\[0.0ex]  
\end{array}$  \\ \cline{2-2}  
  & & &  & \vspace{-0.10in} \\
  & $\begin{array}{c}  
\\
(10,0) \\[-4.0ex] 
\end{array} $
& &  &  \\ \cline{1-1}\cline{3-5} 
  & & &  & \vspace{-0.10in} \\
$\begin{array}{c} 
(1,0) \\ 
\textrm{(Unstable)} \end{array} $ 
& & $\!-0.0001$ & $10^{11}$ & 1.2 hr \\ \hline\hline  
 & &  &  &  \vspace{-0.10in} \\
\hspace{-0.10in} 
$\begin{array}{c} 
\infty
\end{array} $ 
& & \ \ $0.0001$ & $5 \! \times \! 10^{11}$ & 6 hr  \\ 
  &  $\begin{array}{c}  
(-0.35,0) \\[-2.0ex] 
\end{array} $
 & &  &  \vspace{-0.12in} \\
\cline{1-1}\cline{3-5} 
 & &  &  \vspace{-0.10in} \\
$\begin{array}{c} 
\textrm{The most outer LC} \\ 
\textrm{of the 3 small LCs} \\ 
\textrm{(Unstable)} \\[-2.0ex]  
\end{array} $ 
 &  & 
$\begin{array}{c} 
-0.001 \\[-2.5ex]  
\end{array}$ 
& 
$\begin{array}{c} 
10^{12} \\[-2.5ex]  
\end{array}$ 
& 
$\begin{array}{c} 
12 \ \textrm{hr} \\[-2.5ex]  
\end{array}$ 
\\ \cline{2-2} 
& $\begin{array}{c} 
(-0.25,0) \\[-8.0ex]  
\end{array} $
 & &  &   \\ \cline{1-1}\cline{3-5}  
  & & &  & \vspace{-0.10in} \\
$\begin{array}{c}  
\\[-5.0ex] 
\textrm{The middle LC} \\ 
\textrm{of the 3 small LCs} \\ 
\textrm{(Stable)} \\[-10.0ex]  
\end{array} $
&  & 
$\begin{array}{c} 
\\ 
\ \ 0.01 \\[-5.0ex]  
\end{array} $
 & 
$\begin{array}{c} 
\\[0.0ex] 
2 \! \times \! 10^{11} \\[-5.0ex]  
\end{array} $
& 
$\begin{array}{c} 
\\[0.0ex]  
25 \ \textrm{hr} \\[-5.0ex]  
\end{array} $
\\  
  & & &  &  \vspace{-0.10in} \\
& & & & \\ \cline{2-2} 
  & & &  &  \vspace{-0.10in} \\
& $\begin{array}{c} 
\\ 
(-0.05,0) \\[-4.0ex]  
\end{array} $
 & &  &   \\ \cline{1-1}\cline{3-5}  
  & & &  & \vspace{-0.10in} \\
$\begin{array}{c} 
\\[-5.0ex]  
\textrm{The most inner LC} \\ 
\textrm{of the 3 small LCs} \\ 
\textrm{(Unstable)} \\[-10.0ex]  
\end{array} $
&  & 
$\begin{array}{c} 
-0.02 \\[-8.0ex]  
\end{array} $
 & $ 
\begin{array}{c}  
8 \! \times \! 10^{12}  \\[-4.0ex]  
\end{array} 
$  & 
$ 
\begin{array}{c}  
100 \ \textrm{hr} \\[-4.0ex] 
\end{array} 
$ \\  
  & & &  &  \vspace{-0.10in} \\
& & & & \\ \cline{2-2} \cline{4-5} 
  & & & $
\begin{array}{c}  
10^{13} \\[-4.ex] 
\end{array}$ & 
$ 
\begin{array}{c} 
125 \ \textrm{hr} \\[-4.0ex] 
\end{array} $  \\
& $\begin{array}{c} 
\\ 
(-0.002,0) \\[-6.0ex]  
\end{array} $
 & & & \\ \cline{1-1}\cline{3-5} 
  & & &  &  \vspace{-0.10in} \\
$\begin{array}{c}  
\\[-4.5ex] 
(0,0) \\ 
\textrm{(Stable)} \\[-5.0ex]  
\end{array} $
& & 
$\begin{array}{c} 
\\[-1.0ex]  
\ \ \, 0.02 \\[-2.0ex]  
\end{array} $
& 
$\begin{array}{c} 
\\[-1.0ex]  
7 \! \times \! 10^{12} \\[-2.0ex]  
\end{array} $
& 
$\begin{array}{c} 
\\[-1.0ex]  
87.5 \ \textrm{hr} \\[-2.0ex]  
\end{array} $
\\  
  & & &  &  \vspace{-0.10in} \\
  & & &  &  \\ 
\hline
\end{tabular}
} 
\vspace{0.20in} 
\end{table}

\section{Conclusion} 
In this paper, we have considered bifurcation of limit cycles in 
planar quadratic systems and numerically simulate four limit cycles 
which are visualizable. After we study the two well-known 
classical examples, and a recently published system, we focus 
on near-integrable systems, and construct a concrete example to show 
visualizable four limit cycles.

\section*{Acknowledgments}
This work was supported by the Natural Sciences and Engineering 
Research Council of Canada (No.~R2686A02). Useful discussions with 
Drs. Mathieu Desroches Jean-Pierre Françoise and Junmin Yang 
are greatly appreciated.


\end{document}